\documentclass[12pt]{iopart}
\usepackage{graphicx}
\usepackage{verbatim}
\usepackage{url}
\usepackage{geometry}
\usepackage{lscape}
\usepackage{subfigure}
\begin{document}
\title{A multi-scale model for correlation in B cell VDJ usage of zebrafish}
\author{Keyao Pan$^1$ and Michael W Deem$^{1,2}$}
\address{$^1$ Department of Bioengineering, Rice University, 6100 Main Street, Houston, TX 77005, USA}
\address{$^2$ Department of Physics and Astronomy, Rice University, 6100 Main Street, Houston, TX 77005, USA}
\ead{mwdeem@rice.edu}
\begin{abstract}
The zebrafish (\emph{Danio rerio}) is one of the model animals for study of immunology because the dynamics in the adaptive immune system of zebrafish are similar to that in higher animals. In this work, we built a multi-scale model to simulate the dynamics of B cells in the primary and secondary immune responses of zebrafish.  We use this model to explain the reported correlation between VDJ usage of B cell repertoires in individual zebrafish. We use a delay ordinary differential equation (ODE) system to model the immune responses in the 6-month lifespan of a zebrafish.  This mean field theory gives the number of high affinity B cells as a function of time during an infection.  The sequences of those B cells are then taken from a distribution calculated by a ``microscopic'' random energy model.
This  generalized $NK$ model shows that mature B cells specific to one antigen largely possess a single VDJ recombination.  The model allows first-principles calculation of the probability, $p$, that two zebrafish responding to the same antigen will select the same VDJ recombination.
This probability $p$ increases with the B cell population size and the B cell selection intensity.  The probability $p$ decreases with the B cell hypermutation rate. The multi-scale model predicts correlations in the immune system of the zebrafish that are highly similar to that from experiment.
\end{abstract}
\submitto{\PB}
\maketitle

\section{Introduction}

B cell-mediated adaptive immunity exists in jawed animals \cite{Matsunaga1998}. B cells protect hosts by secreting antibodies that recognize and neutralize pathogens and foreign substances. Immunity generated by B cells is hence indispensable to the hosts' survival. The primary immune response occurs when a novel type of antigen is detected by the immune system. The antigen is processed and presented to na\"ive B cells, which mature in the germinal center. In the maturation process, the B cells acquire the capability to recognize and neutralize a specific antigen. First, a na\"ive B cell recombines one V gene segment, one D gene segment, and one J gene segment in the genome to create the nucleotide sequence encoding the antibody. Second, this nucleotide sequence undergoes multiple rounds of somatic hypermutation, and B cells with high affinity to the antigen are selected. The selected mature B cells differentiate into the antibody-secreting plasma cells or long-lived memory B cells that  activate the secondary immune response against the same antigen in the future.
We will use the generalized $NK$ model, discussed in more detail below, to describe this maturation process.
Janeway \emph{et al}.\ provide a more detailed review of the B cell-mediated immune reaction \cite{Janeway2005}. Understanding the dynamics of and relationship between VDJ recombination and somatic hypermutation informs one about the central mechanism of B cell immunity.

Recent experimental studies provide information on the B cell maturation process in zebrafish. Zebrafish (\emph{Danio rerio}) have been increasingly used as a model animal to study the immune system because experiments on zebrafish are easy to perform, zebrafish reproduce quickly, and zebrafish possess one of the most primitive adaptive immune systems, which is a model for the adaptive immune systems in humans and mice \cite{Trede2004,Thisse2002,Yoder2002}. The genome of zebrafish contains 39 V gene segments, five D gene segments, and five J gene segments, which together encode the V region of immunoglobulin IgM heavy chain in zebrafish \cite{Danilova2000,Danilova2005}. High-throughput sequencing of the complete IgM repertoires in 14 six-month-old zebrafish revealed that one fish carries up to 5000--6000 distinct nucleotide sequence of IgM with $39 \times 5 \times 5 = 975$ VDJ recombinations \cite{Weinstein2009}. VDJ usage, the probability that each of the 975 possible VDJ recombinations is used in the IgM repertoire, has a correlation coefficient between individual zebrafish up to $r = 0.75$ \cite{Weinstein2009}.

We seek to explain these experimental data from reasonably first principles.
We use delay ordinary differential equations (ODEs) to describe the immune response of a zebrafish as it encounters
antigens over its lifetime.  The mean field number of B cells predicted by the ODE model are then assigned sequences based upon the results of the more ``microscopic'' generalized $NK$ (GNK) model.
The  generalized $NK$ model is used to analyze correlations in different zebrafish exposed to similar antigens.
The model describes the dynamics and the localization of the VDJ recombination in primary and secondary immune responses.
The immune response in a single fish usually localizes upon a single VDJ recombinant, with some hypermutational diversity in sequences that are progeny of the VDJ recombination.
We use the model presented here  to calculate the probability, $p$, that two fish localized on the same VDJ recombination.

For our calculations, we use the GNK model as it has been derived in the literature.
The original $NK$ model was
constructed to describe evolution of short peptide fragments in phage display experiments
 \cite{Kauffman1987,Kauffman1989}.
The GNK model was introduced to model protein evolution experiments \cite{Bogarad1999}.
In particular, it accounts for the formation of and interaction between secondary structures of a protein and for the presence of an active or binding site in the protein.  The GNK model was used to validate the performance of a new, hierarchical protocol for protein
molecular evolution.
 We will be using the GNK model without change of parameter values, and so calculations in the present work are predictions.
The GNK model was used to compute the expected efficacy of B cell vaccines, as a function of the difference between the vaccine and infecting virus \cite{Deem2003}.
The GNK model was generalized to T cell immunity and parametrized on altered peptide ligand experiments \cite{Park2004}.
Averages and correlations in the immune response were predicted.  The GNK model was used to look at autoimmune
disease \cite{Sun2005}.
The GNK model was applied to T cell immunodominance data, and the predictions were within the error bars of human clinical
trial data \cite{Zhou2006}.
The GNK model was used to predict influenza H3N2 vaccine efficacy, and the predictions were within 10\% of human efficacy data
\cite{Gupta2006}.
The GNK model was used to examine the T cell response to cancer \cite{Yang2006}.
A mechanism was put forward to explain immunodominance in cancer vaccine studies, and a quantitative comparison
of GNK model predictions to specific lysis data was made.
The GNK model was further used to predict tumor escape and immune elimination probabilities as a function of expressed epitopes and vaccination strategy \cite{Yang2006b}.
It was shown that  allele loss is more significant than point mutation for tumor escape.
The GNK model was used to predict the usefulness of reduced alphabets in protein evolution experiments \cite{Munoz2008}.
The GNK model was used to predict immunity in an agent-based influenza epidemiology model, predicting epidemic progression in accordance with WHO data
\cite{Zhou2009}.
The theory was used to  predict the impact of vaccination strategy, time of vaccination, and extent of vaccination on epidemic progression.
Finally, a review of the GNK model, with a table of parameter values, can be found in \cite{Sun2006} and
\cite{Zaman2009}.
A discussion of the theory behind the spin-glass approach to biological evolution can be found in \cite{Lorenz2011}.
Mora \emph{et al}.\ fit a random energy model, similar to the $NK$ model, with a large number of parameters, approximately $10^3$, to experimentally measured probabilities of D gene segment usage in zebrafish \cite{Mora2010}.
We here use the much smaller subset of parameters that have been determined in the GNK model.
%KEYAO FIX REFS

The purpose of this study is to model the B cell-mediated immune response and to explain the correlations in the VDJ usage between pairs of zebrafish. The theory presented here is used to describe the mechanism of B cell-mediated immunity and to depict a snapshot of B cell repertoires at any time point in the host's life span. The Materials and Methods section describes the ODE model as it is applied to multiple types of antigen and the generalized $NK$ model as it is applied to  a single type of antigen. The Results and Discussion section compares the predictions to experimental data. Finally, we present our conclusions and outlook.

\section{Materials and methods}

\subsection{Motivation and description of the multi-scale model}

We seek to explain Weinstein et al.'s data \cite{Weinstein2009} from first principles. We develop a model to predict the B cell-mediated immune response in a  zebrafish living in an environment with multiple types of antigen. 
We also predict the correlations in the immune response between a pair of such zebrafish. The immune response of zebrafish contains quantities covering many orders of magnitude. First, the zebrafish were raised for six months in the experiment \cite{Weinstein2009}. On the other hand, the duration of a primary or secondary immune response is  about 10 days \cite{Janeway2005}. The number of B cells in a zebrafish is on the order  of $10^5$ \cite{Weinstein2009}, although only a small fraction of these B cells react to a given antigen \cite{Janeway2005}.

\begin{figure}[!ht]
\begin{center}
\includegraphics[width=6in]{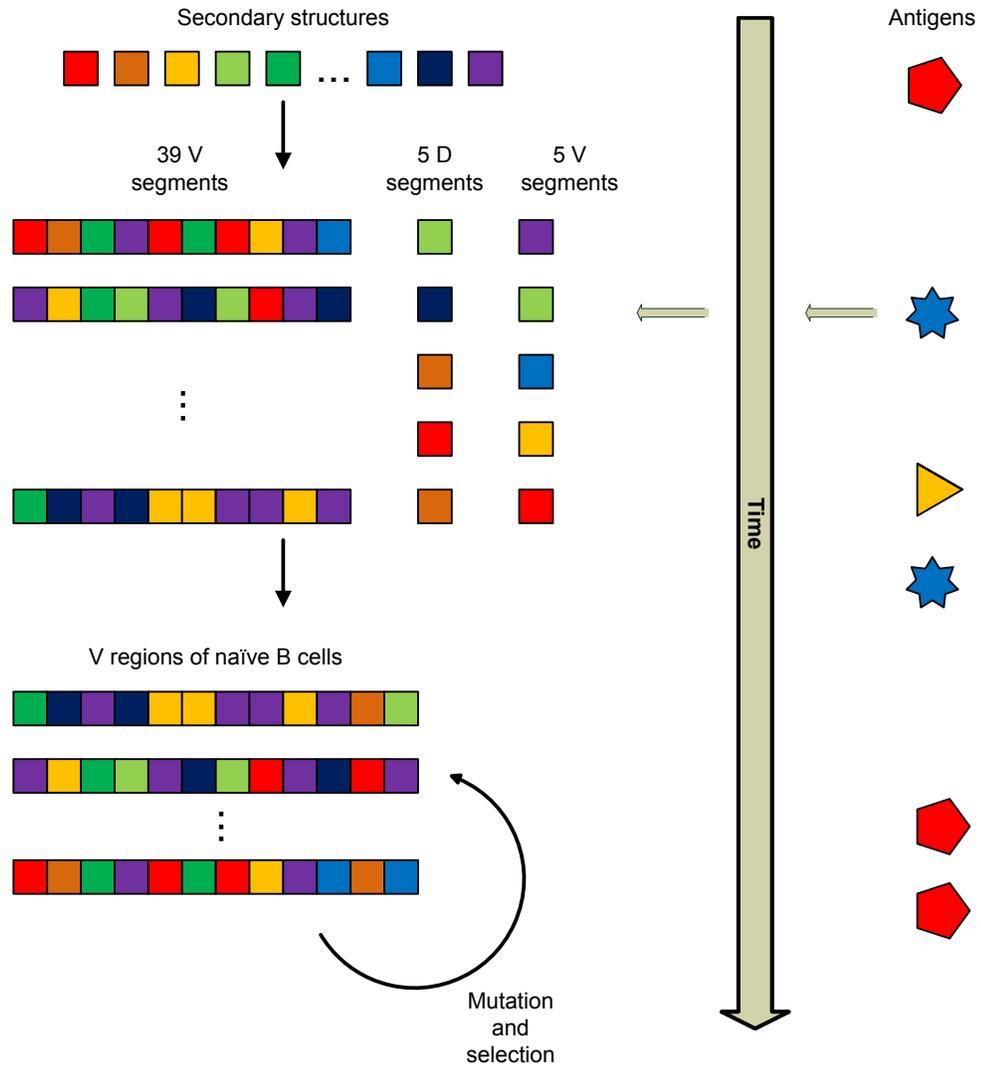}
\end{center}
\caption{Illustration of the multi-scale model of zebrafish immune response. The timeline arrow represents the zebrafish life history from 0 (hatch) to 180 days. To the right of the timeline is the delay ODE model. The zebrafish are challenged by antigen at several random time points. Each challenge leads to a primary or secondary immune response. To the left of the timeline arrow is the generalized $NK$ model described by a flow chart. Distinct secondary structures are randomly recombined to form V, D, and J segments, which randomly recombined to form the variable region of the IgM heavy chain. The variable region underwent 30 rounds of mutation and selection in the primary immune response and another 30 rounds in the secondary immune response. In this figure, as an example, the generalized $NK$ model describes the primary immune response against the second antigen, represented by the blue star,  this zebrafish met in its lifetime.}
\label{fig:scheme}
\end{figure}

 Due to this wide range of scales, we use multi-scale modeling to describe these data.
 The delay ODE system computes the dynamics of the number of mature B cells in the zebrafish during exposure to antigen. The delay ODE gives the average number of plasma cells, $N_{\mathrm{p}i}$, and memory B cells, $N_{\mathrm{m}i}$, specific to antigen $i$ as a function of time. Note that $N_{\mathrm{p}i}$ is identical in a pair zebrafish, and so is $N_{\mathrm{m}i}$ in this mean field theory. The generalized $NK$ model computes the immune responses in a pair of zebrafish against an antigen. One result from the generalized $NK$ model is that due to selection, the VDJ usage in one fish after the primary response is almost always localized to a single VDJ recombination. The generalized $NK$ model computes the probability, $p$, that this VDJ recombination is the same in a pair of fish.
 The output of the generalized $NK$ model, $p$, is used to assign the sequences of the B cells computed in the delay ODE model.
We consider a pair of zebrafish, because we will compute pairwise correlations for comparison to the experimental data.  We assign the same VDJ sequence to the pair with probability $p$, otherwise we assign a different VDJ sequence to each fish.
 Table \ref{tab:model} provides the comparison between these two models. Figure \ref{fig:scheme} illustrates these two models.

\subsection{ODE model}
\label{sec:ode_model}

The delay ordinary differential equation (ODE) system computes the dynamics of the immune response triggered by antigens. The ODE system computes the average, mean-field number of B cells responding to antigen as a function of time.  The number of mature B cells in a zebrafish can reach the order of magnitude of $10^3$ \cite{Weinstein2009}. Zebrafish lived in an environment with $N_\mathrm{ag}$ types of antigens. For one zebrafish, inoculation of antigen $i$ triggers the immune response that boosts the number of relatively short-lived B cells and long-lived memory B cells in a zebrafish.   The immune response of each zebrafish against antigen $i$ is modeled by two delay ODEs:
\begin{eqnarray}
\label{eq:ode_1}
\frac{\mathrm{d}x_i\left(t\right)}{\mathrm{d}t} & = c_1 v_i\left( t-\tau_1 \right) + c_3 v_i\left(t\right) y_i\left(t\right) - b x_i\left(t\right) \qquad & i = 1, 2, \dots, N_\mathrm{ag}\\
\label{eq:ode_2}
\frac{\mathrm{d}y_i\left(t\right)}{\mathrm{d}t} & = c_2 v_i\left( t-\tau_2 \right) \qquad & i = 1, 2, \dots, N_\mathrm{ag}.
\end{eqnarray}
in which state variables $x_i$ and $y_i$ are the numbers of antigen $i$ specific plasma cells, which secret antibodies, and memory B cells, respectively. The initial conditions at the time of hatching are $x_i \left(0\right) = 0$, $y_i \left(0\right) = 0$ since hatchlings are assumed not to have seen antigen. The level of antigen $i$ received by the zebrafish, $v_i$, is the result of a random process. Antigen inoculation is random, but the same when a pair of zebrafish is considered, since the environment is common for both zebrafish. We assume that $N_\mathrm{ag} = 10$ distinct types of antigen exist in the environment. The zebrafish are in one of two states, the normal state in which the antigen is absent, and the infected state in which the antigen is present in the zebrafish. A newborn zebrafish is not inoculated by antigens and so it is in the normal state, $v_i\left(0\right) = 0$. During antigen inoculation, both zebrafish receive one randomly selected type of antigen, denoted antigen $i$. The average time span between antigen inoculations  is $\lambda = 30$ days. The value of $v_i\left(t\right)$ jumps from 0 to 100 at antigen inoculation, and the zebrafish transit from the normal state to the infected state. The antigen presentation lasts for approximate one week \cite{Berek1991}. Therefore, the zebrafish transit from the infected state to the normal state, and $v_i$ falls back to zero after seven days if the host was not inoculated by antigen $i$ in the past seven days. The zebrafish stay in the infected state if it is reinnoculated with antigen during the seven days. The ODE system described above contains $2N_\mathrm{ag}$ equations for a pair of zebrafish.

When the host receives antigen $i$ at time $t$, a primary immune response is triggered if the host is na\"ive to this antigen ($y_i \left(t\right) = 0$). Otherwise a secondary immune response is triggered. Antigen $i$ stimulates production of antigen $i$ specific plasma cells with rate $c_1 v_i \left(t-\tau_1\right)$ and memory B cells with rate $c_2 v_i \left(t-\tau_2\right)$, according
to Eqs.\ (\ref{eq:ode_1}) and (\ref{eq:ode_2}).
 In the primary immune response, B cells appear around $\tau_1=5$ days after the initial contact of antigen, and memory B cells appear around $\tau_2=30$ days after the contact \cite{Janeway2005}. In the secondary immune response, existing memory B cells mount an immediate reaction to the corresponding antigen with a rate $c_3 v_i \left(t\right) y_i \left(t\right)$, which is higher than the rate $c_1 v_i \left(t-\tau_1\right)$ in the primary immune response because $y_i \left(t\right) \gg 1$ . The first order term $bx_i$ in equation \ref{eq:ode_1} quantifies the decay of antigen $i$ specific plasma cells with rate $b = 5 \mbox{ day}^{-1}$, because most B cells in germinal centers live a short time before apoptosis \cite{Janeway2005}.  We set s $c_1 = 1$, $c_2 =0.1$, and $c_3 = 0.03$.

We solve the ODE system by numerical integration. The solution gives the composition of mature B cells in a zebrafish challenged by multiple types of antigen during its life history. The   na\"ive repertoire plus the mature B cells at $t = 180$ days represent the B cell repertoire of a 6-month-old zebrafish.

\subsection{Generalized $NK$ model}

The generalized $NK$ model defines the rugged random potential energy landscape, which defines the B cell fitness for the evolutionary process as it depends on the antibody variable region \cite{Bogarad1999,Deem2003,Earl2004,Park2004,Sun2007,Munoz2008}. The generalized $NK$ model assigns a secondary structure and energy to each antibody variable region.  Sequences undergo somatic hypermutation, and selection causes the lower energy sequences to be propagated \cite{Bogarad1999}.
The energy of an antibody variable region is expressed by
\begin{equation}\label{eq:Utotal}
U = \sum_{i=1}^M U_{\alpha_i}^\mathrm{sd} + \sum_{i>j=1}^M U_{ij}^\mathrm{sd-sd} + \sum_{i=1}^P U_i^\mathrm{c}
\end{equation}
in which $M$ is the number of secondary structural subdomains, $P$ is the number of amino acids contributing to the antibody-antigen binding process. The energy $U$ is the sum of three components: secondary structural subdomain energies ($U_{\alpha_i}^\mathrm{sd}$), subdomain-subdomain interaction energies ($\sum_{i>j=1}^M U_{ij}^\mathrm{sd-sd}$), and chemical binding energies ($\sum_{i=1}^P U_i^\mathrm{c}$). The parameters of the generalized $NK$ model are fixed in our previous papers \cite{Bogarad1999,Deem2003,Sun2006}
and \cite{Zaman2009}.
%KEYAO FIX REFS
The secondary structural subdomain energy is expressed by
\begin{equation}\label{eq:Usd}
U_{\alpha_i}^\mathrm{sd} = \sqrt{\frac{1}{M\left(N-K+1\right)}} \sum_{j=1}^{N-K+1} \sigma_{\alpha_i}\left(a_j,a_{j+1},\dots,a_{j+K-1}\right)
\end{equation}
in which $N$ is the number of amino acids in a subdomain and each amino acid in the subdomain interact with $K-1$ other amino acids in the same subdomain. Here $N = 10$ and $K = 4$. An antibody V region contains approximately 120 amino acids \cite{Janeway2005} or equivalently $M = 12$ secondary structures. The identity of the subdomain is denoted by $\alpha_i$. There are $L = 5$ types of subdomains, which are helices, strands, loops, turns, and others, Therefore $\alpha_i = 1, \dots, 5$. The identity of amino acid in position $j$ is represented by $a_j$. The 20 amino acids are categorized into $Q = 5$ classes, which are negative, positive, polar, hydrophobic, and other. So $a_j=1, \dots, 5$. For each of $L$ types of subdomains, $\sigma_{\alpha_i}$ is a $K$-dimensional quenched Gaussian array with zero mean and unit standard deviation.

The subdomain-subdomain interaction energy is expressed by
\begin{equation}\label{eq:Usdsd}
U_{ij}^\mathrm{sd-sd} = \sqrt{\frac{2}{DM\left(M-1\right)}} \sum_{k=1}^D \sigma_{ij}^k \left(a_{j_1}^i, \dots, a_{j_{K/2}}^i; a_{j_{K/2+1}}^j, \dots, a_{j_K}^j\right)
\end{equation}
Between subdomains $i$ and subdomain $j$ ($i < j$), $D = 6$ interactions occurs. Each interaction involves $K/2$ interacting amino acids in subdomain $i$ and $K/2$ interacting amino acids in subdomain $j$. For each pair of subdomains $\left(i, j\right)$, $\sigma_{ij}^k$ is a $K$-dimensional quenched Gaussian array with zero mean and unit standard deviation.

The chemical binding energy is expressed by
\begin{equation}\label{eq:Uc}
U_i^\mathrm{c} = \sqrt{\frac{1}{P}} \sigma_i\left(a_i\right)
\end{equation}
in which $a_i$ is the identity of one of the $P = 5$ amino acids in the antibody variable region contributing to the binding to the antigen. The quantity $\sigma_i \left(a_i\right)$ is a quenched Gaussian with zero mean and unit standard deviation.

The generalized $NK$ model is used to compute the VDJ distribution in a single fish and in a pair of zebrafish.
We set the number of B cells responding to a particular antigen as
 $N_\mathrm{size} = 2000$ because the number of B cells in a germinal center is in the order of magnitude of $10^3$ \cite{Jacob1991,Smith1997}. These VDJ recombinants undergo evolution and selection \cite{Gupta2006}.
%KEYAO FIX
The variable region contains approximately 120 amino acids in which 100, 10, and 10 amino acids are encoded by the V, D, and J gene segment, respectively. Thus the generalized $NK$ model fixes the length of variable region to 120 amino acids and defines amino acid 1--100, 101--110, and 111--120 as encoded by the V, D, and J gene segment, respectively.
 By randomly selecting and recombining VDJ fragments  \cite{Bogarad1999}, we created 39 V segments consisting of 10 secondary structures, five D segments consisting of one secondary structure, and five J segments consisting of one secondary structure.
The initial $N_\mathrm{size}=2000$ structures
are created by recombination from these V, D, and J segments.
These sequences then undergo rounds of somatic hypermutation.
The mutation rate in this process is roughly $n_\mathrm{mut} = 0.5$ amino acid/structure/generation \cite{Janeway2005}; thus we set the number of mutated amino acid in each generation in each structure to follow a Poisson distribution with mean $\lambda=0.5$. Cutoff selection is used, with $p_\mathrm{cut} = 20\%$ of the structures with the lowest energy  propagated to the next generation \cite{Deem2003}. The primary immune response lasts about 10 days, or equivalently 30 generations of B cells \cite{Berek1991,Janeway2005}; thus we apply the generalized $NK$ model for 30 rounds of somatic hypermutation and selection. A secondary immune response is another 30 rounds of somatic hypermutation. The distribution of VDJ usage in each zebrafish and in a pair of zebrafish is computed.

The general scheme of the model is shown in figure \ref{fig:scheme}.

\subsection{Model usage}

The ODE system models the number of antigen $i$ specific B cells in each zebrafish in a mean-field approach. The duration of primary and secondary immune responses, which are explicitly modeled by the number of iterations in the generalized $NK$ model, are respectively treated with the duration $\tau_1$ and $\tau_2$ in the ODE system. The distribution of VDJ usage in the generalized $NK$ model
is used to assign sequences to each of these B cells.

The predictions for correlations in VDJ usage between a pair of fish from this model of the immune response were compared to experiment \cite{Weinstein2009}.The high-throughput sequencing experiment by Weinstein \emph{et al}.\ measured the IgM heavy chain frequencies of 975 VDJ recombination in each zebrafish.  The experiment also measured the correlation in VDJ between all pairs of zebrafish.
 The experimentally measured  correlation coefficients were classified into four bins: independence (correlation coefficient $r \le 0.1$), low correlation ($0.1 < r \le 0.2$), moderate correlation ($0.2 < r \le 0.5$), and high correlation ($r > 0.5$).  This histogram of the correlation coefficients will be one of the predictions of our model.

\section{Results and discussion}

\subsection{Overview of the results}
\label{sec:overview}

In  subsection \ref{sec:response} we present the results of the generalized $NK$ model for the distribution and correlation of VDJ genotypes.
Subsection \ref{sec:sensitivity} presents a sensitivity analysis to the number of na\"ive B cells reacting to the antigen, the somatic hypermutation rate, and the selection intensity. In subsection \ref{sec:du}, the distribution of energy changes $\Delta U$ of single mutations in the generalized $NK$ model is presented.
The results of the delay ODE model in response to random antigen challenge are presented in subsection \ref{sec:ode}. Combining the results of the two models, we present the correlation coefficient between the VDJ usage in a pair of zebrafish.
We will show that the predicted distribution of correlation coefficients of VDJ usage in a pair of fish matches
the experimentally one \cite{Weinstein2009} ($p\mbox{-value} = 0.62$, Pearson $\chi^2$ test).

\subsection{Primary and secondary immune response against one antigen}
\label{sec:response}

\begin{figure}[!ht]
\begin{center}
\subfigure[]{
\includegraphics[width=4in]{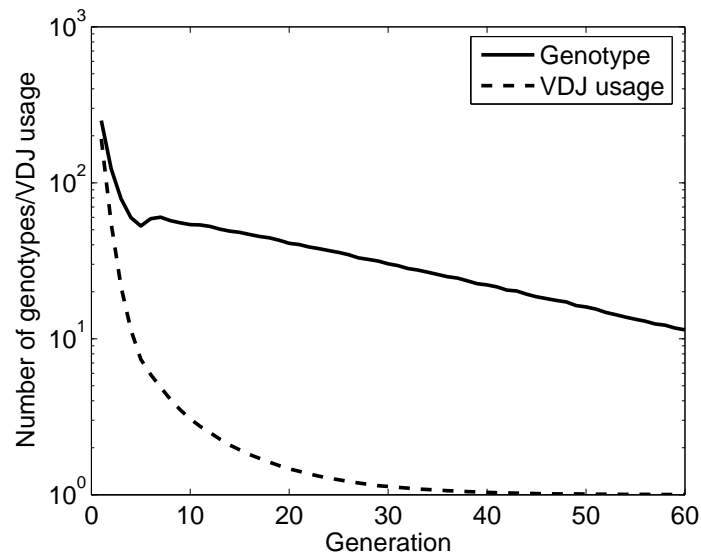}
\label{fig:n_vdj_a}
}
\subfigure[]{
\includegraphics[width=4in]{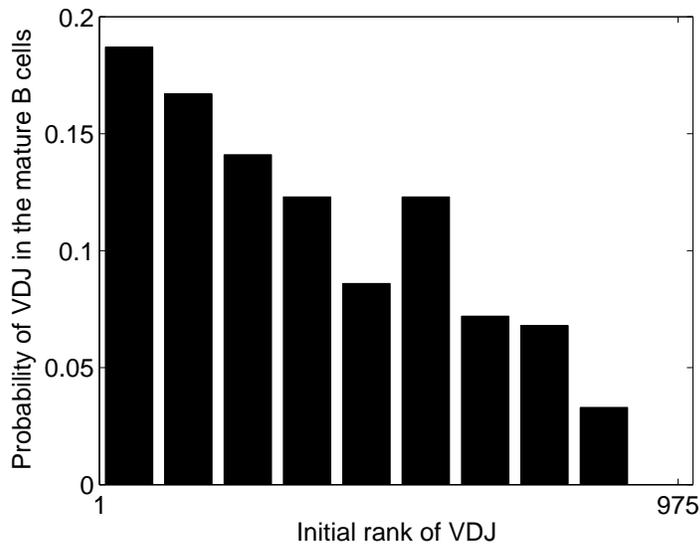}
\label{fig:n_vdj_hist}
}
\end{center}
\caption{\subref{fig:n_vdj_a} The numbers of distinct genotypes and VDJ recombinations in the primary (generation 1--30) and secondary (generation 31--60) immune responses against one antigen involving $N_\mathrm{size} = 2000$ na\"ive B cells. The number of VDJ recombinations decreased much faster than that of B cell genotypes. In most cases all the B cells showed 1--2 VDJ recombinations at the end of the primary immune response and one VDJ recombination at the end of the secondary immune response. \subref{fig:n_vdj_hist} Probability distribution at generation 60 of the rank of probabilities of the $39 \times 5 \times 5 = 975$ VDJ recombinations in $N_\mathrm{size} = 2000$ na\"ive B cells reacting to one type of antigen. This probability distribution used 1000 rank data generated by running the generalized model 1000 times. The na\"ive VDJ ranks fell into 10 bins with rank 1--100, 101--200, $\dots$, 901--975. Bin 10 with rank 901--975 was empty.}
\label{fig:n_vdj}
\end{figure}

A distribution of immune responses is calculated in the generalized $NK$ model.
The diversity of VDJ recombinants decreases as the immune response proceeds, as shown in figure \ref{fig:n_vdj}\subref{fig:n_vdj_a}. Initially,  the average number of  VDJ recombinations was 849.6. During the primary immune response, the average number of genotypes rapidly decreased to 30.2 at generation 30 in an exponential way. In the secondary immune response, the average number of genotypes decreased from the value of 30.2 to 11.4 at generation 60 following the same exponential function of time. As shown in figure \ref{fig:n_vdj}\subref{fig:n_vdj_a}, localization in the VDJ recombination space occurred prior to that in the sequence space. Most of the sequence diversity in a single fish at the end of the primary response is the result of a single initial VDJ recombinant.  The average number of VDJ recombinations reduced from the initial value of 849.6 to 1.13 at generation 30 at the end of the primary immune response, and further to 1.003 at generation 60 at the end of the secondary immune response. Thus, the B cells converged to one dominant VDJ recombination in nearly all immune responses. A dominant VDJ recombination exists in the 2000 B cells at the end of both the primary and the secondary immune response. In a sample of 1000 runs, 950 runs have the fraction of sequences that match the dominant one greater than 0.85 at the end of the primary immune response, and over 950 runs have the fraction equal to 1 at the end of the secondary immune response. The dominant VDJ recombination in two different zebrafish is identical at the end of the primary and the secondary immune responses with probability $p_1 = 0.326$ and $p_2 = 0.327$, respectively. Because $p_1$ and $p_2$ are similar, we defined $p=0.327$ as the probability that two different zebrafish have the same dominant VDJ recombination at the end of an antigen-specific immune response, whether primary or secondary.

We calculate the probability with which an initially highly fit VDJ recombinant becomes the dominant VDJ recombination in the immune response.
Figure \ref{fig:n_vdj}\subref{fig:n_vdj_hist} shows how the probability of becoming dominant depends on the initial fitness ranking.  VDJ recombinations with higher ranks tend to be eventually fixed.  This correlation of fitness with fixation is what creates correlation in the VDJ usages between a pair of fish that was observed in experiment \cite{Weinstein2009}.

\begin{figure}[!ht]
\begin{center}
\includegraphics[width=4in]{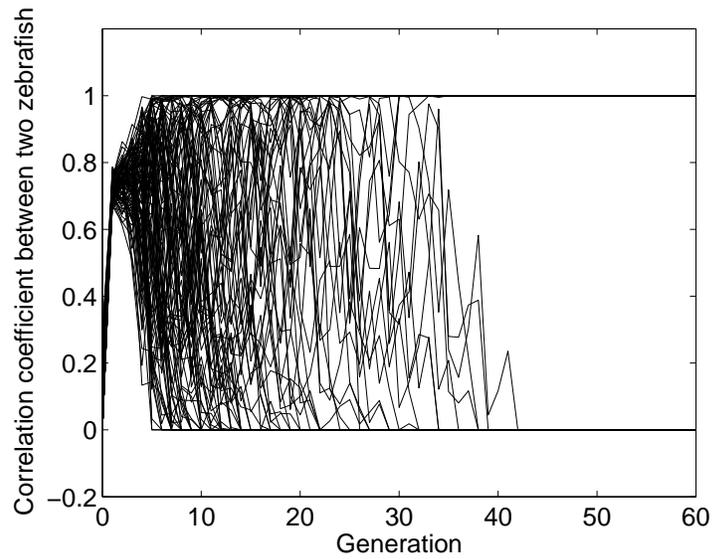}
\end{center}
\caption{Trajectories of correlation coefficients, $r$, between VDJ usage of B cells in two zebrafish reacting to one certain antigen. The simulation consisted of 1000 runs, each of which generated the correlation coefficient, $r$, between na\"ive B cells in generation 0 and their progeny in generations 1--30 in the primary immune response and in generations 31--60 in the secondary immune response. The first 100 out of 1000 trajectories are here plotted for clarity.}
\label{fig:corrcoef}
\end{figure}

We now calculate the correlation in VDJ usage in a pair of zebrafish infected with the same antigen.  As described in figure \ref{fig:corrcoef}, na\"ive B cells at generation 0 in a pair of zebrafish show uncorrelated VDJ usage. The correlation coefficient increases rapidly to around 0.5 in the first three generations and after that showed large variation in different pairs. At the end of the primary immune reaction with 30 generations, most mature B cells in a single fish are evolved from the same VDJ recombination.  The probability for the correlation in a pair of fish to be $r > 0.995$ at the end of the primary and secondary immune responses is 0.301 and 0.327, respectively. The probability for $r < 0.1$ at the end of the primary and secondary immune responses is 0.641 and 0.672, respectively.

These results show that each VDJ recombination initially in the na\"ive repertoire expands in a local neighborhood by somatic
hypermutation.  Eventually, only one of these clouds exists in the final mature repertoire.
The initial energy is  modestly correlated with whether the VDJ recombinant will survive in the mature repertoire.
The trajectories of correlation coefficients in figure \ref{fig:corrcoef} converge to $-1/975$ or 1 because the mature B cells against one antigen in most individuals show only one VDJ recombination, see figure \ref{fig:n_vdj}\subref{fig:n_vdj_a}. The correlation coefficient between VDJ usage in two zebrafish is $-1/975$ if the mature B cells have distinct VDJ recombinations in the two zebrafish. The correlation coefficient is 1 if the VDJ recombination is identical in the two zebrafish. Correlations $-1/975$ and 1 are, therefore, two absorbing states in the random process of correlation coefficients in the immune response. Figure \ref{fig:corrcoef}  shows that the VDJ usage in B cells is localized by the end of the primary immune response; the secondary immune responses does not change the VDJ usage in most cases.

\subsection{Parameter sensitivity in the generalized $NK$ model}
\label{sec:sensitivity}

\begin{figure}[!ht]
\begin{center}
\subfigure[]{
\includegraphics[width=2.5in]{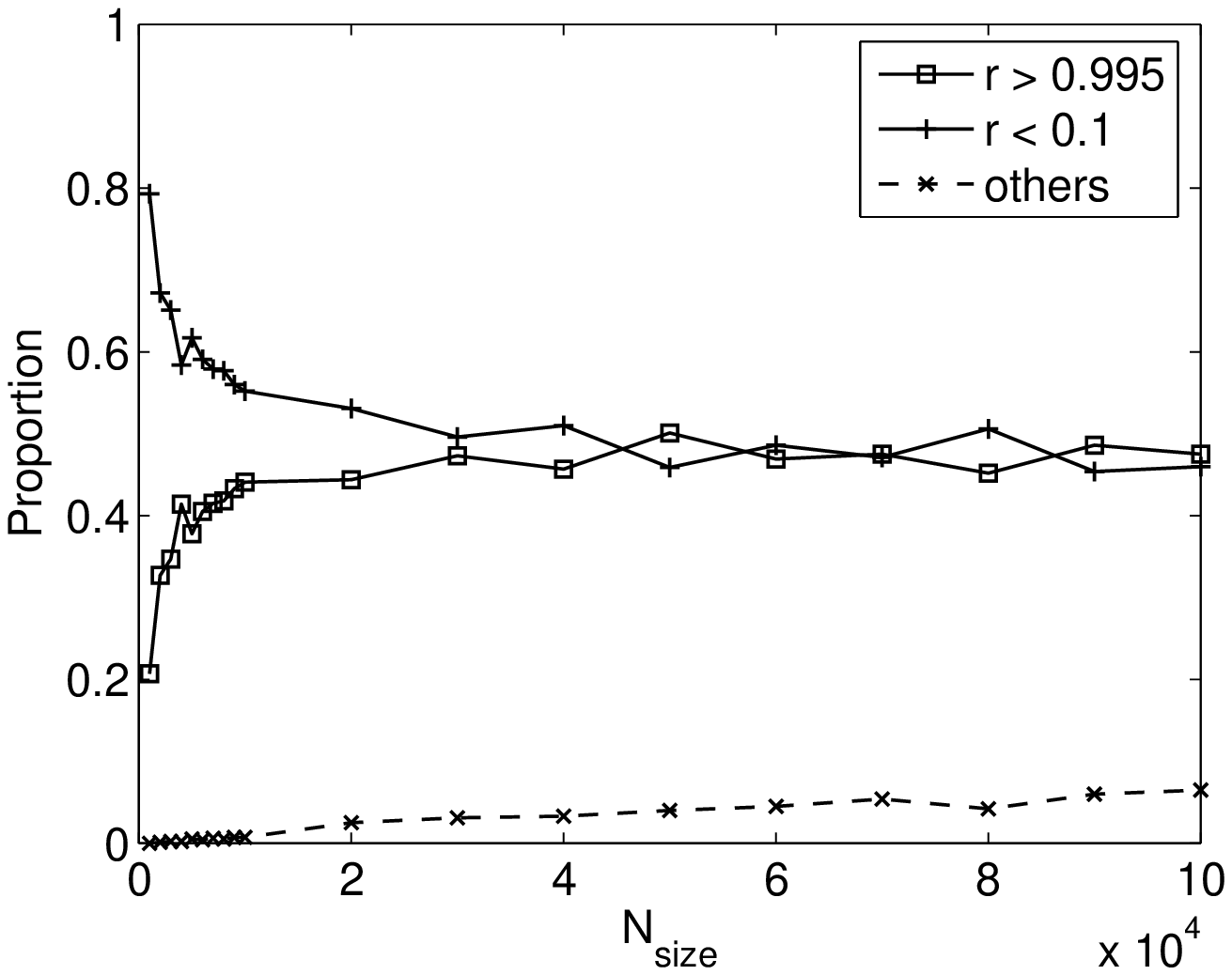}
\label{fig:N_size}
}
\subfigure[]{
\includegraphics[width=2.5in]{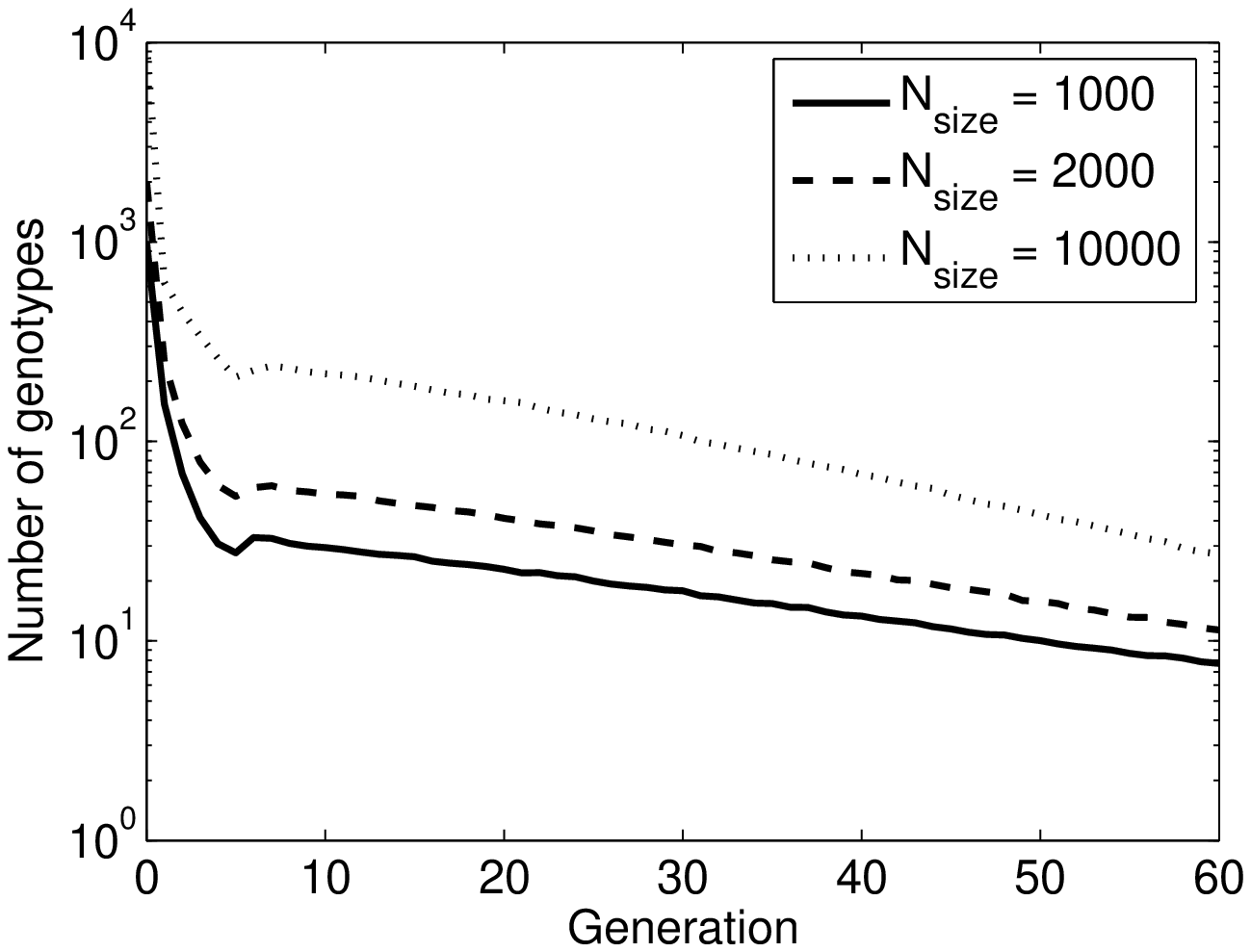}
\label{fig:N_size_N_seq}
}
\subfigure[]{
\includegraphics[width=2.5in]{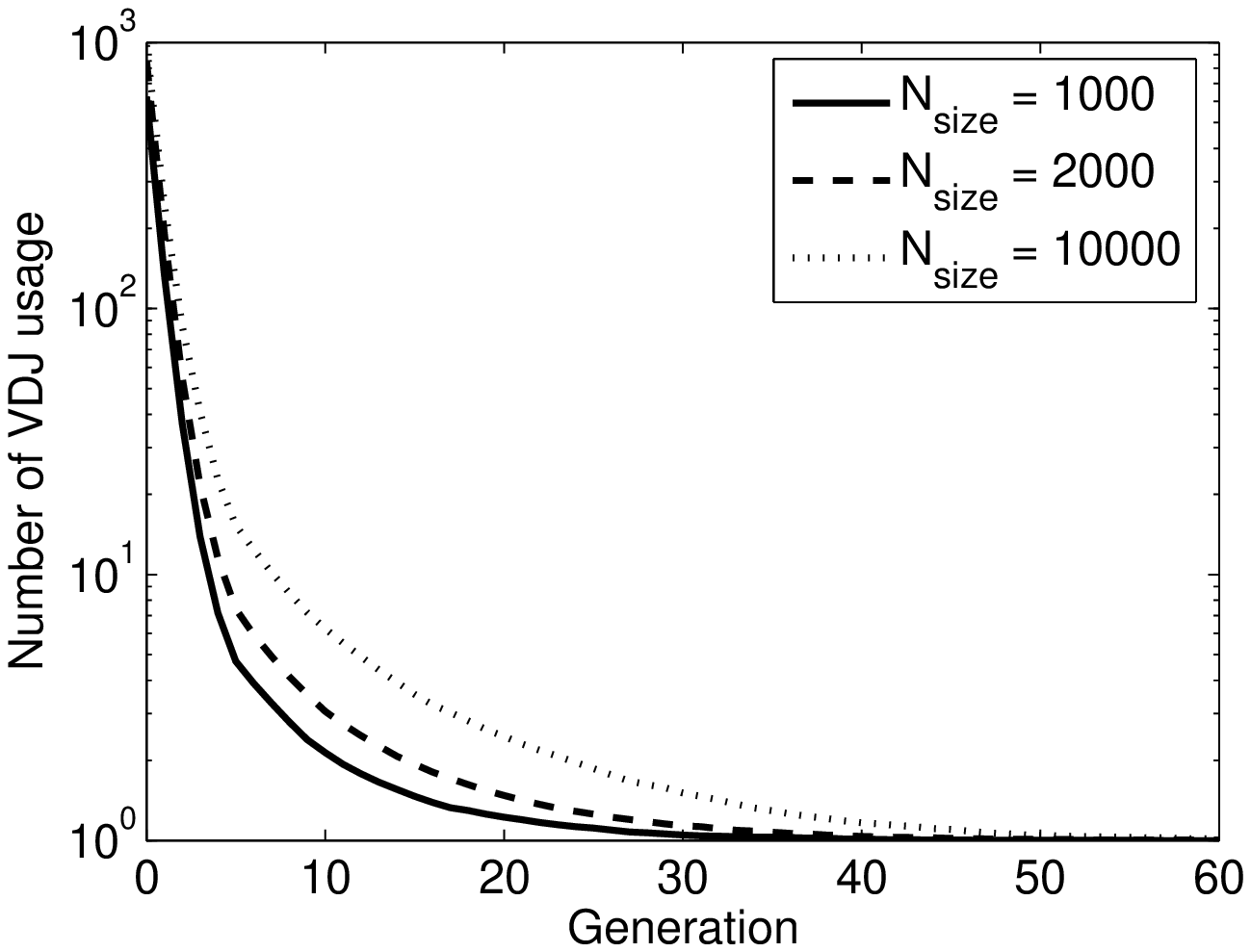}
\label{fig:N_size_N_vdj}
}
\end{center}
\caption{\subref{fig:N_size} Relationship between the number of B cells reacting to one antigen, defined as $N_\mathrm{size}$, and the correlation data between two zebrafish. Measured by the generalized $NK$ model, the correlation coefficient, $r$, between the VDJ usage in the mature B cells from the secondary immune response fell into three categories: identical ($r > 0.995$), distinct ($r < 0.1$), and unfixed VDJ usage, with probabilities $p$, $q$, and $1 - p - q$, respectively. The values of $p$, $q$, and $1-p-q$ are plotted respectively as the functions of $N_\mathrm{size}$ ranging from $10^3$ to $10^5$. \subref{fig:N_size_N_seq} The number of distinct genotypes in each generation of the B cells in primary immune response. The $x$- and $y$- axes are the same as those in figure \ref{fig:n_vdj}\subref{fig:n_vdj_a}. This diagram presents the dynamics of genotype numbers in three cases, $N_\mathrm{size}=1000$, $N_\mathrm{size}=2000$, and $N_\mathrm{size}=10000$, respectively. \subref{fig:N_size_N_vdj} Same as \subref{fig:N_size_N_seq}, except for the number of different VDJ recombinations.}
\label{fig:N_size_analysis}
\end{figure}

As shown in section \ref{sec:response}, B cells in a pair of zebrafish converge to the same VDJ recombination with probability $p$ and converged to two different VDJ recombinations with probability $q$. There exists a small probability $1-p-q$ that greater than two VDJ recombinations exist in the pair. Figure \ref{fig:N_size_analysis}\subref{fig:N_size} plots $p$, $q$, and $1-p-q$ as the function of $N_\mathrm{size}$. The value of $p$ increased rapidly from 0.207 when $N_\mathrm{size} = 1000$ to 0.475 when $N_\mathrm{size} = 10000$ and were insensitive to $N_\mathrm{size}$ with $N_\mathrm{size} > 10000$. The probability $1-p-q$ for non-converged VDJ recombinations increased with $N_\mathrm{size}$.
The increasing trends of $p$ and $1-p-q$ come from the delayed localization of the VDJ recombination with larger $N_\mathrm{size}$. As illustrated by figures \ref{fig:N_size_analysis}\subref{fig:N_size_N_seq} and \ref{fig:N_size_analysis}\subref{fig:N_size_N_vdj},
localization in the sequence and recombination space occurs more slowly for larger
 $N_\mathrm{size}$.
 The B cells thus explore the energy landscape associated with the sequence space, around each VDJ recombinations, for a longer period of time for larger $N_\mathrm{size}$. That is, a larger value of $N_\mathrm{size}$  enables the B cells to generate more mutants and to explore a broader subregion of the energy landscape, and fixation to a single VDJ recombination occurs more slowly. For small $N_\mathrm{size}$, the B cells in a pair of zebrafish, both of which explore the energy landscape more intensively, have a higher chance to evolve to the same local minimum in the energy landscape.

\begin{figure}[!ht]
\begin{center}
\subfigure[]{
\includegraphics[width=2.5in]{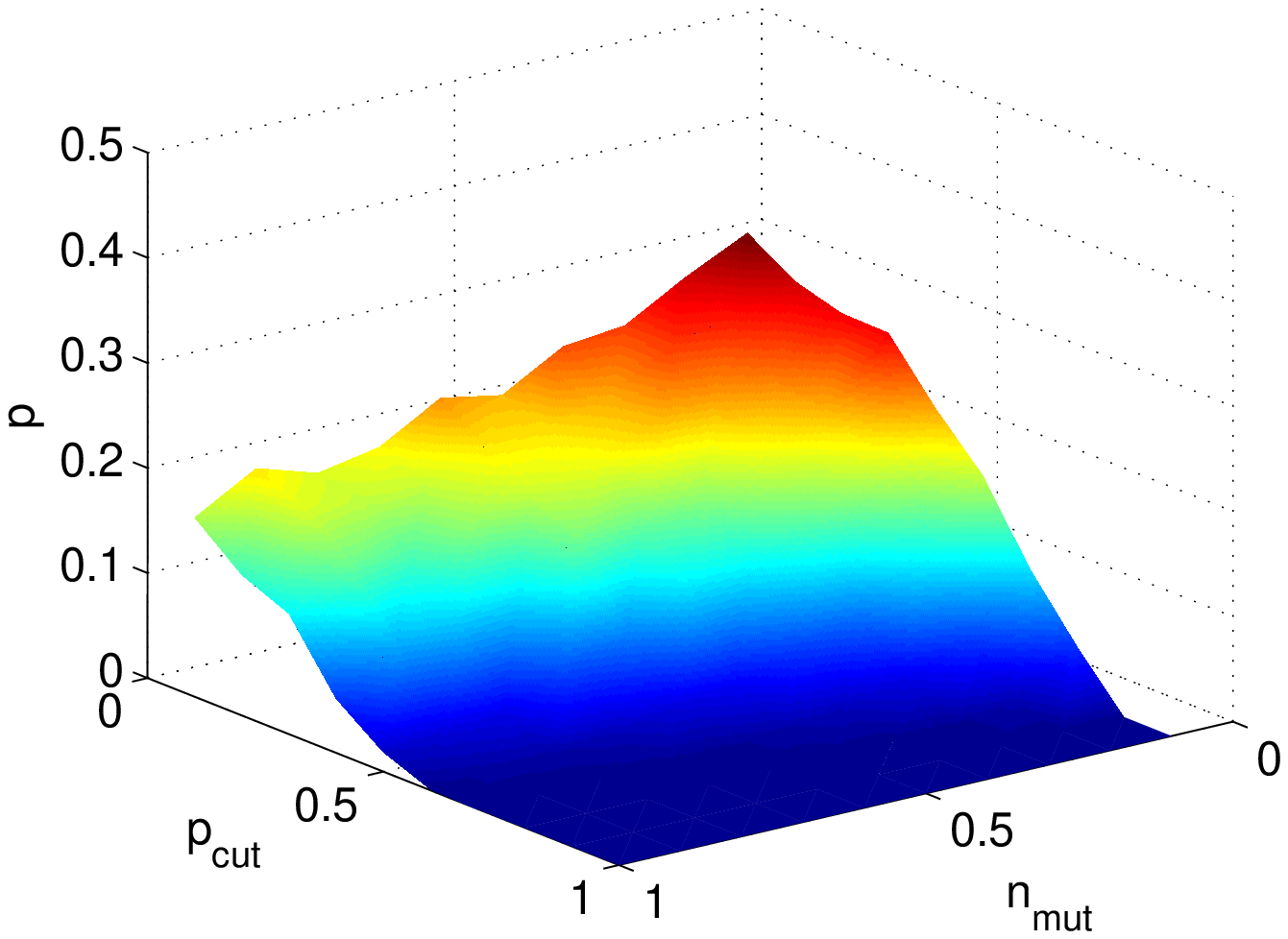}
\label{fig:ps_r2_1000}
}
\subfigure[]{
\includegraphics[width=2.5in]{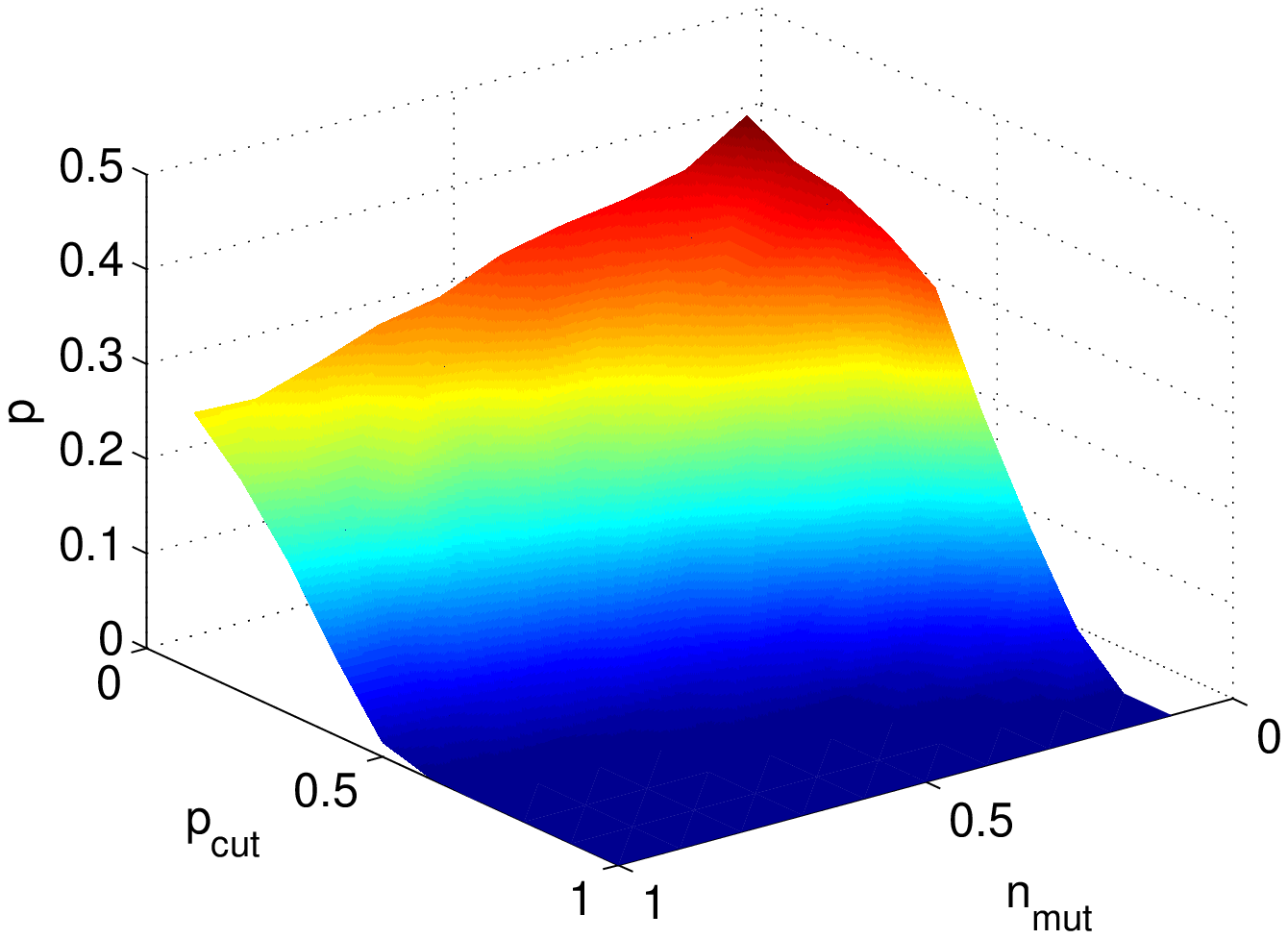}
\label{fig:ps_r2_2000}
}
\subfigure[]{
\includegraphics[width=2.5in]{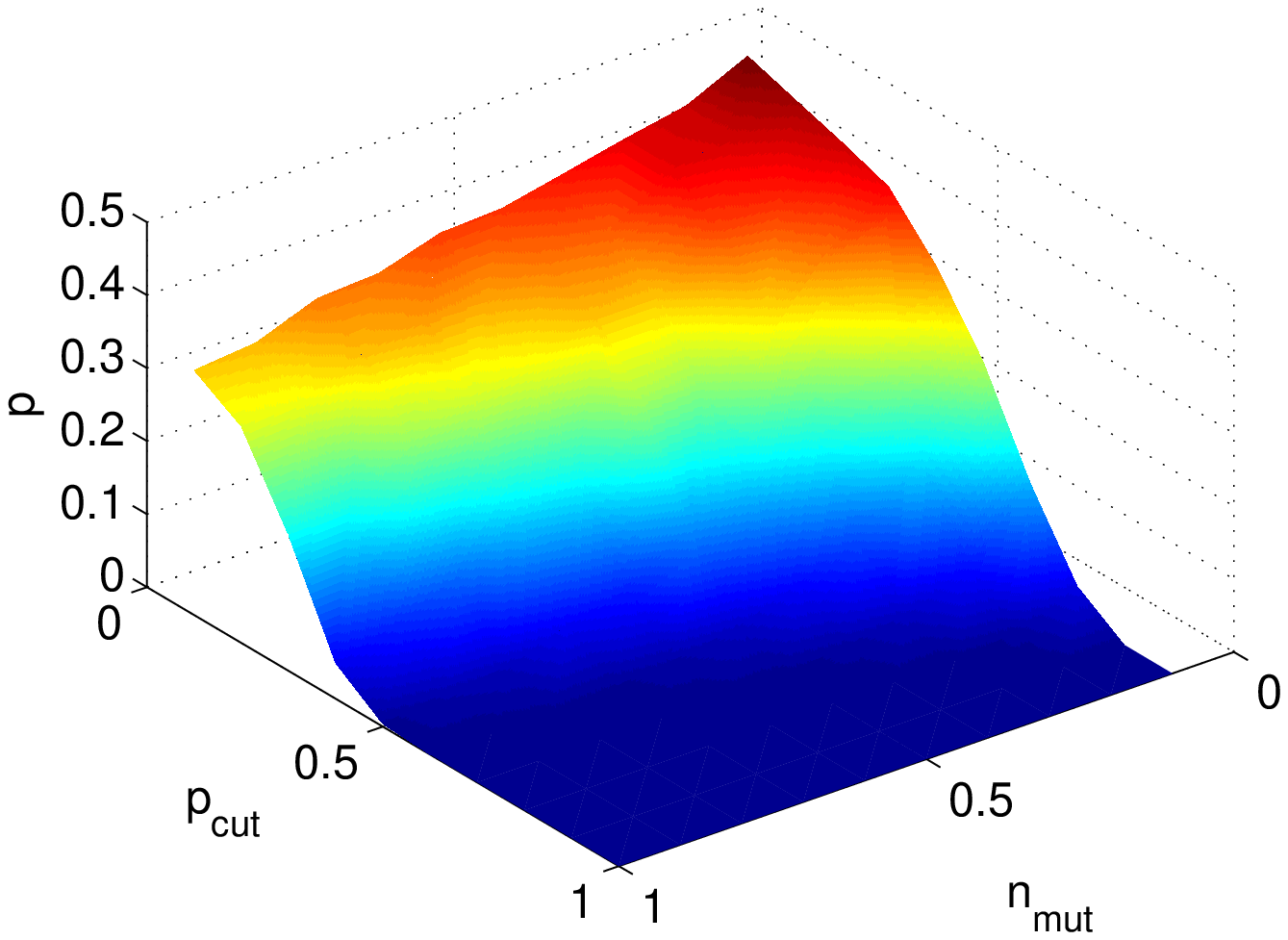}
\label{fig:ps_r2_5000}
}
\subfigure[]{
\includegraphics[width=2.5in]{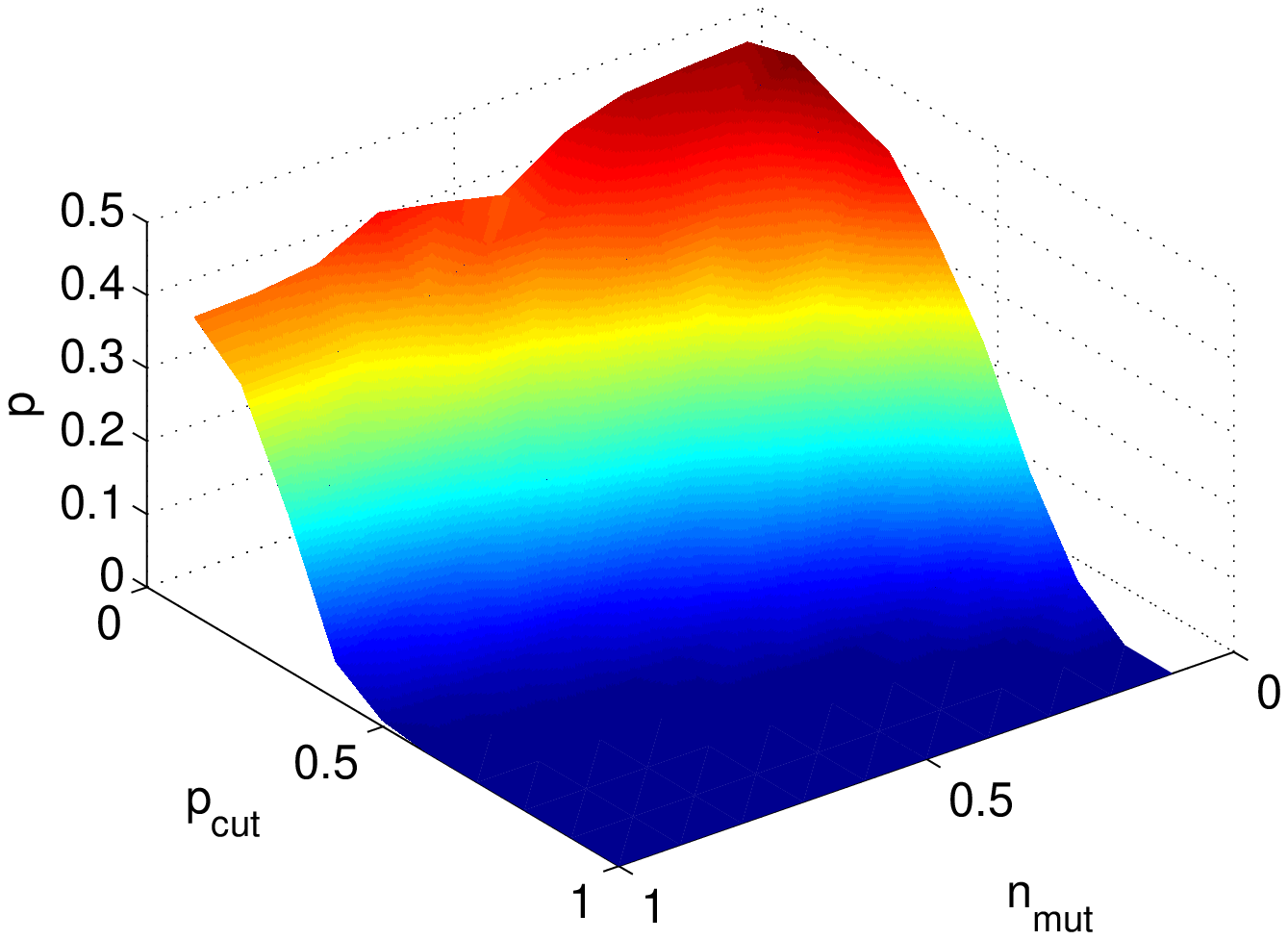}
\label{fig:ps_r2_10000}
}
\end{center}
\caption{The effect of the average number of point mutations in each generation, $n_\mathrm{mut}$, and the proportion of B cells propagated to the next generation, $p_\mathrm{cut}$, on the probability, $p$, that a pair of zebrafish localize upon the same VDJ recombination in the immune response. Each point on the surfaces shows the value of $p$ calculated from the generalized $NK$ model as a function of $n_\mathrm{mut}$ and $p_\mathrm{cut}$. Each of the four sub figures is shown for distinct numbers of antigen-specific B cells $N_\mathrm{size}$: \subref{fig:ps_r2_1000} $N_\mathrm{size}=1000$, \subref{fig:ps_r2_2000} $N_\mathrm{size}=2000$, \subref{fig:ps_r2_5000} $N_\mathrm{size}=5000$, and \subref{fig:ps_r2_10000} $N_\mathrm{size}=10000$.}
\label{fig:ps_r2}
\end{figure}

\begin{figure}[!ht]
\begin{center}
\subfigure[]{
\includegraphics[width=2.5in]{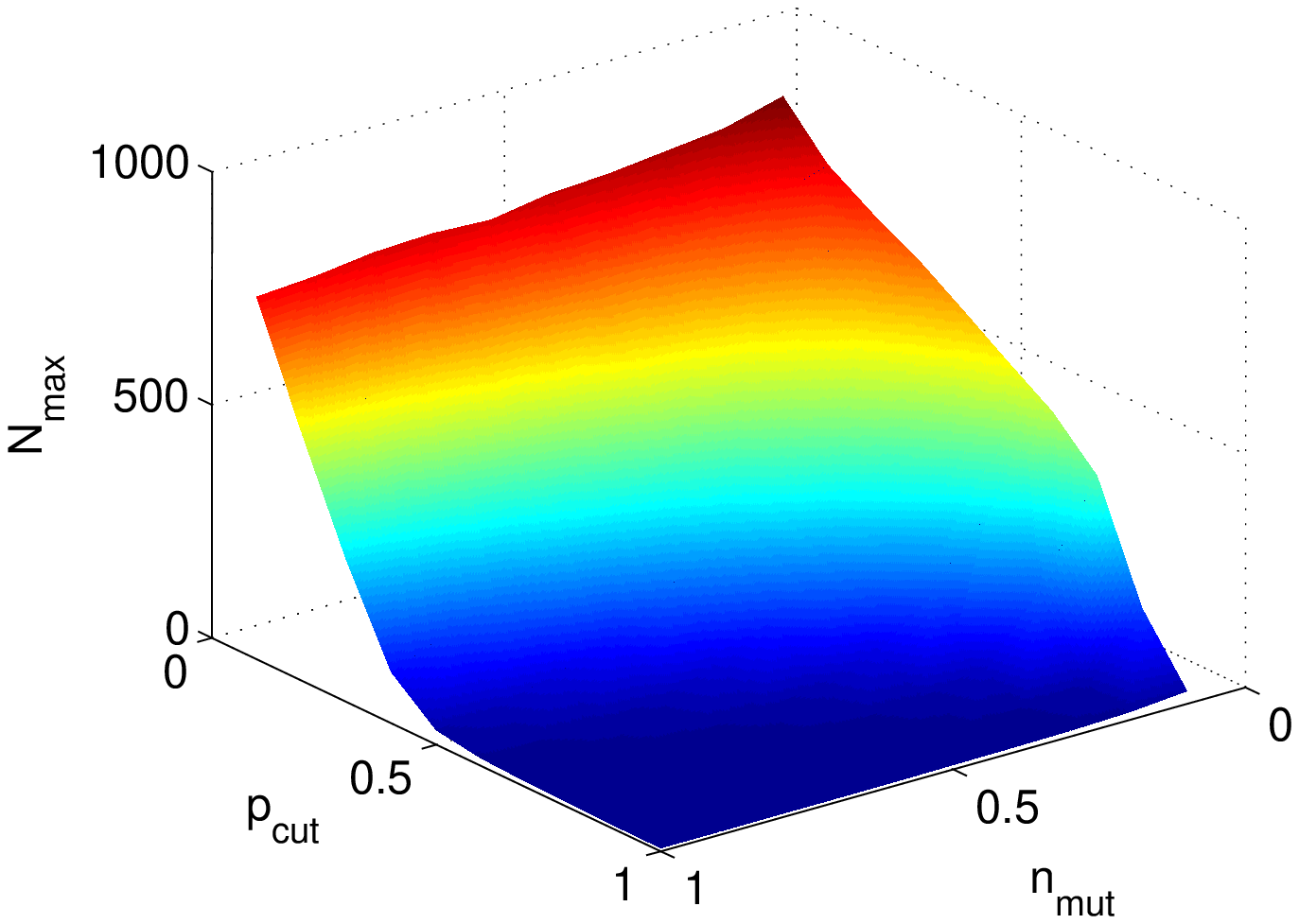}
\label{fig:ps_cluster_1000}
}
\subfigure[]{
\includegraphics[width=2.5in]{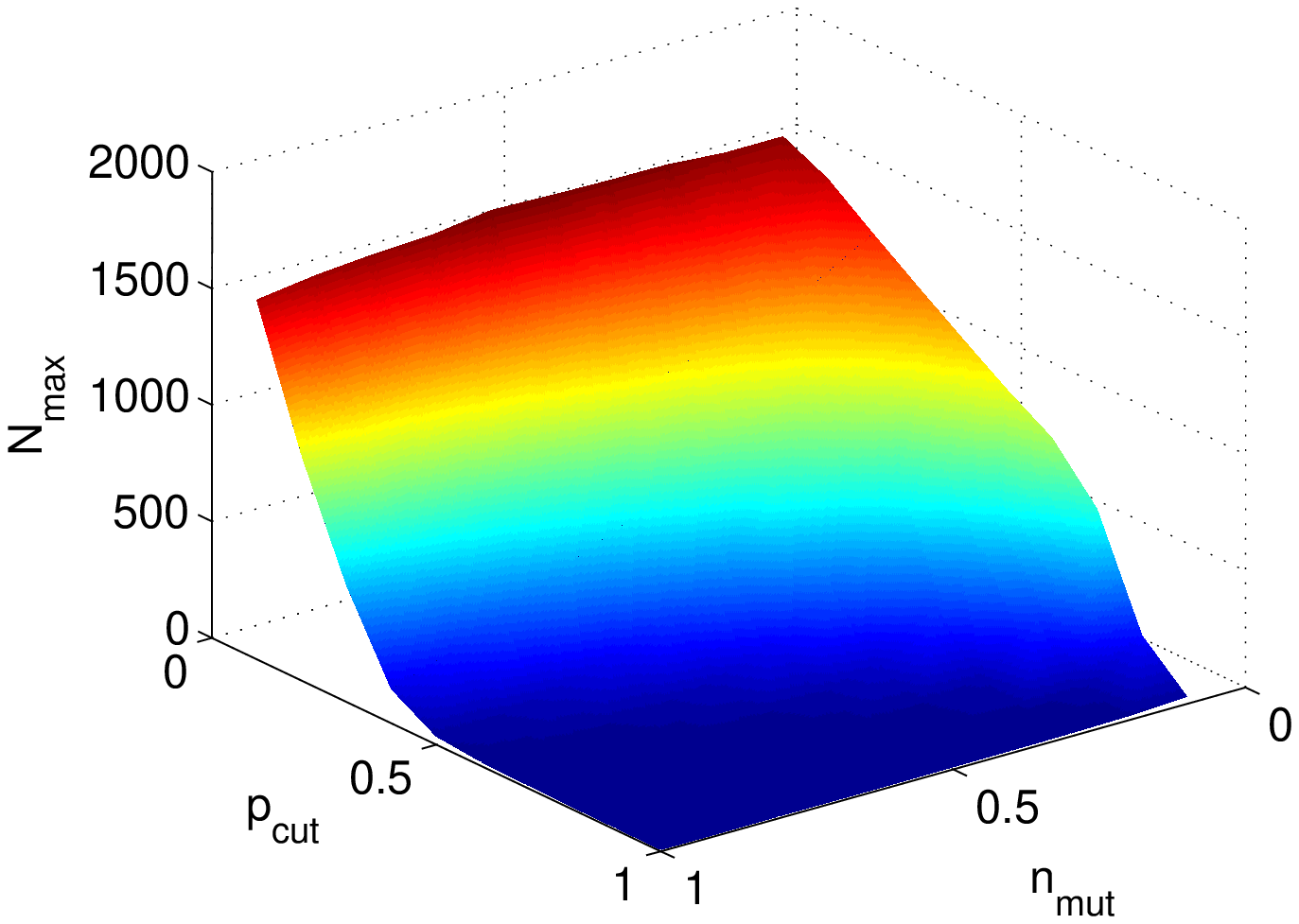}
\label{fig:ps_cluster_2000}
}
\subfigure[]{
\includegraphics[width=2.5in]{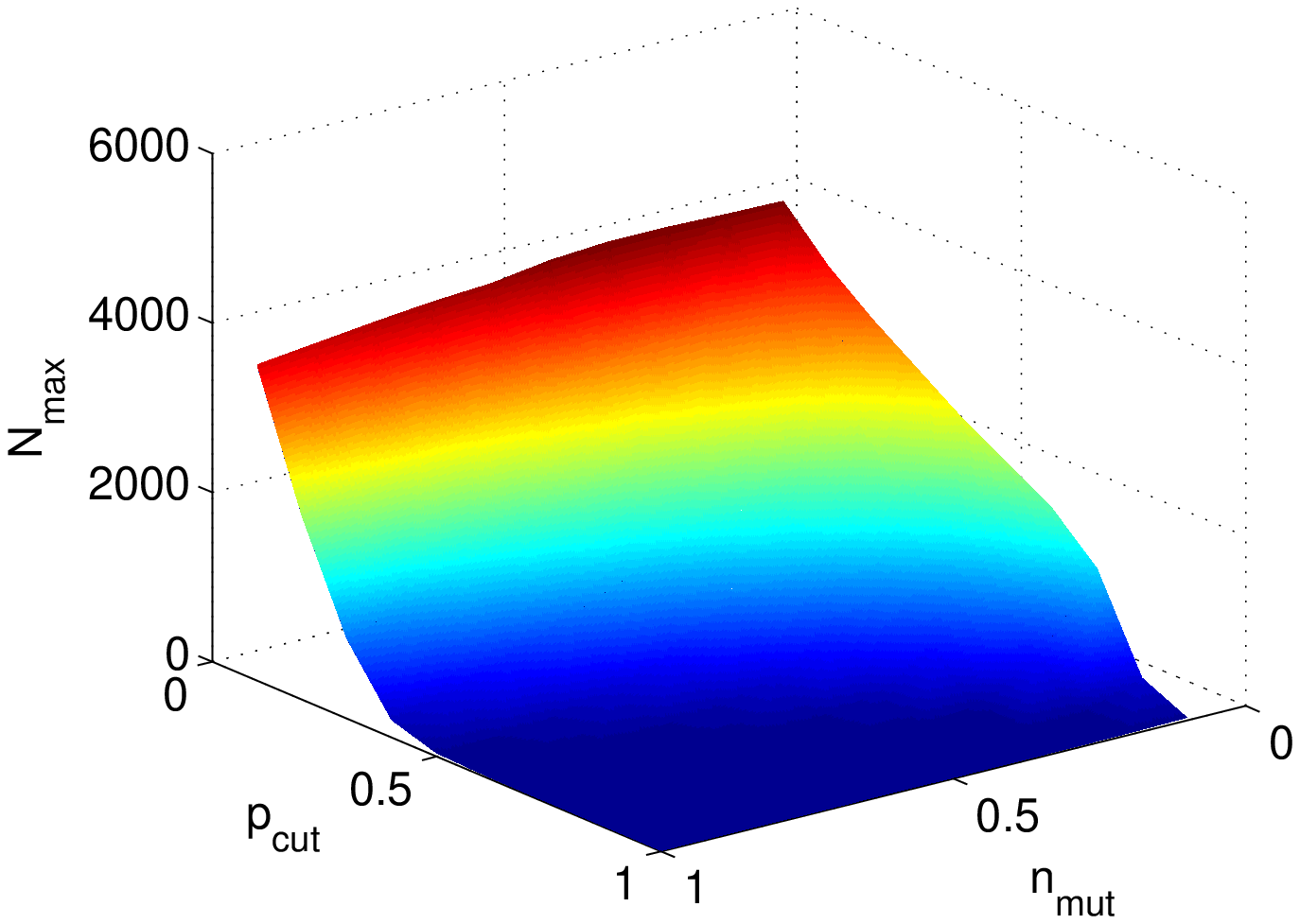}
\label{fig:ps_cluster_5000}
}
\subfigure[]{
\includegraphics[width=2.5in]{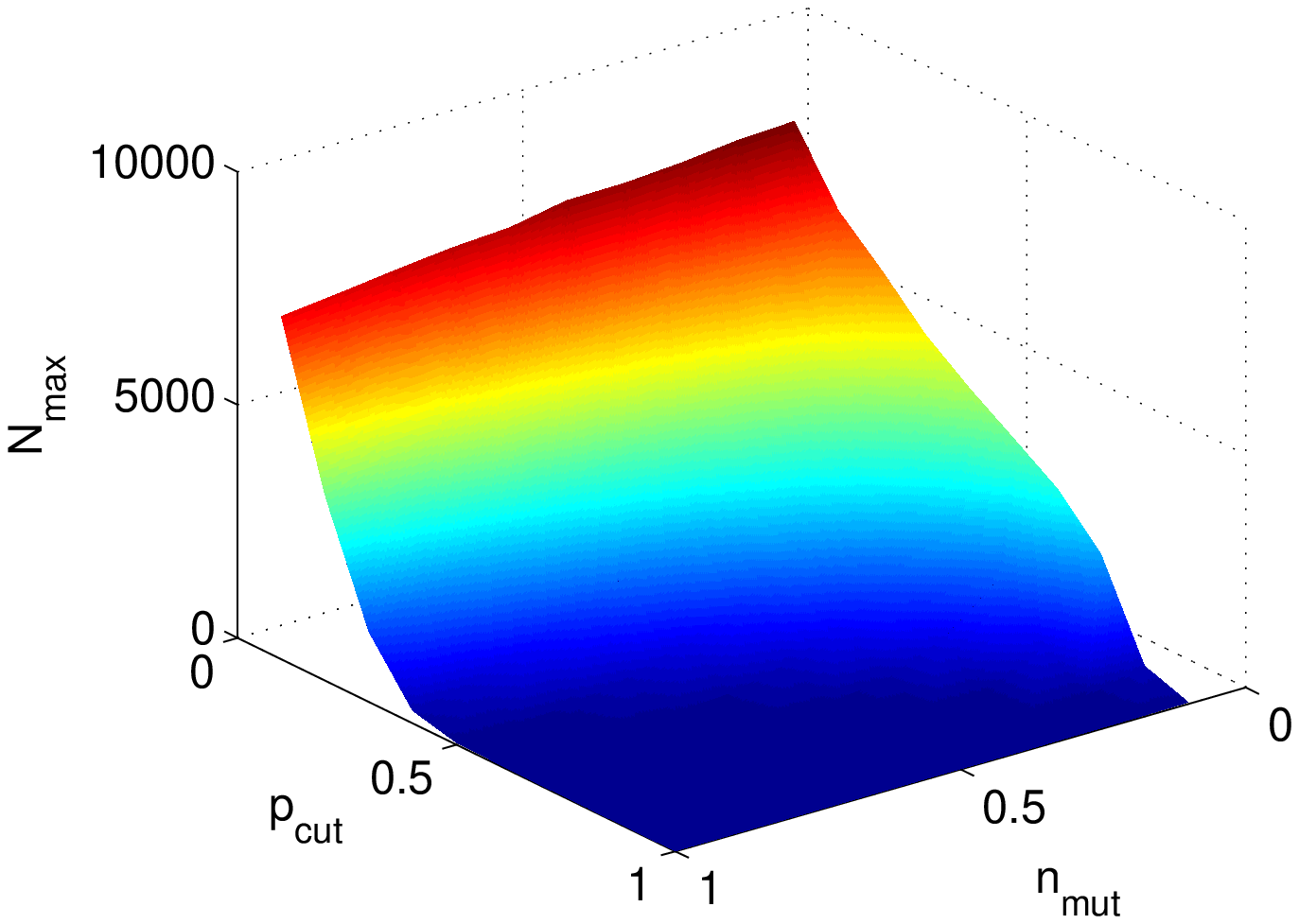}
\label{fig:ps_cluster_10000}
}
\end{center}
\caption{The maximum number of B cells with identical sequence at the end of the secondary immune response, $N_\mathrm{max}$, as a function of $n_\mathrm{mut}$ and $p_\mathrm{cut}$. As in figure \ref{fig:ps_r2}, $N_\mathrm{size}$ as the number of antigen-specific B cells has a constant value in each sub figure: \subref{fig:ps_cluster_1000} $N_\mathrm{size}=1000$, \subref{fig:ps_cluster_2000} $N_\mathrm{size}=2000$, \subref{fig:ps_cluster_5000} $N_\mathrm{size}=5000$, and \subref{fig:ps_cluster_10000} $N_\mathrm{size}=10000$.}
\label{fig:ps_cluster}
\end{figure}

For fixed $N_\mathrm{size}$, the values of $n_\mathrm{mut}$ and $p_\mathrm{cut}$ affect the B cell maturation process. The effect of Both $p$ and $N_\mathrm{max}$, defined as the maximum number of B cells with identical sequence at the end of the primary immune response, depend on $n_\mathrm{mut}$ and $p_\mathrm{cut}$. If $N_\mathrm{max}$ is close to $N_\mathrm{size}$, most of the mature B cells from the primary immune response share the same genotype. The values used in the generalized $NK$ model are $n_\mathrm{mut}=0.5$ and $p_\mathrm{cut}=0.2$. Figures \ref{fig:ps_r2} and \ref{fig:ps_cluster} show the dependence on $n_\mathrm{mut}$ and $p_\mathrm{cut}$. The values of both $p$ and $N_\mathrm{max}$ decrease quickly with $p_\mathrm{cut}$. The values of $p$ and $N_\mathrm{max}$ fall to near zero with $p_\mathrm{cut}>0.9$. Both $p$ and $N_\mathrm{max}$ are less sensitive to $n_\mathrm{mut}$.

The two parameters $n_\mathrm{mut}$ and $p_\mathrm{cut}$ define the mutation and selection in B cell somatic hypermutation. The values of mutation rate, $n_\mathrm{mut}$, and selection pressure, $p_\mathrm{cut}$, observed in experiment have evolved during the evolutionary history of the immune systems \cite{Zhang2010}. The low sensitivity to $n_\mathrm{mut}$ in figures \ref{fig:ps_r2} and \ref{fig:ps_cluster} indicates that with a hypermutation rate $n_\mathrm{mut}$ on the order of magnitude $10^{-1}$/sequence/generation, the immune system maintains the capability to explore the sequence space and to locate the local minima in the energy landscape. On the other hand, the selection pressure $p_\mathrm{cut}$ controls the B cell exploration in the rugged energy landscape. A small value of $p_\mathrm{cut}$ confines the B cells in compact regions in the sequence space by only allowing B cell genotypes with lowest energy to propagate to the next generation and causing premature localization. Smaller values of $p_\mathrm{cut}$ cause both zebrafish to have a higher chance to reach the same local energy minima, as described in figure \ref{fig:ps_cluster}, and result in increased correlation in the VDJ usage, as illustrated in figure \ref{fig:ps_r2}.

\subsection{Distribution of the energy change, $\Delta U$, due to a point mutation}
\label{sec:du}

\begin{figure}[!ht]
\begin{center}
\subfigure[]{
\includegraphics[width=4in]{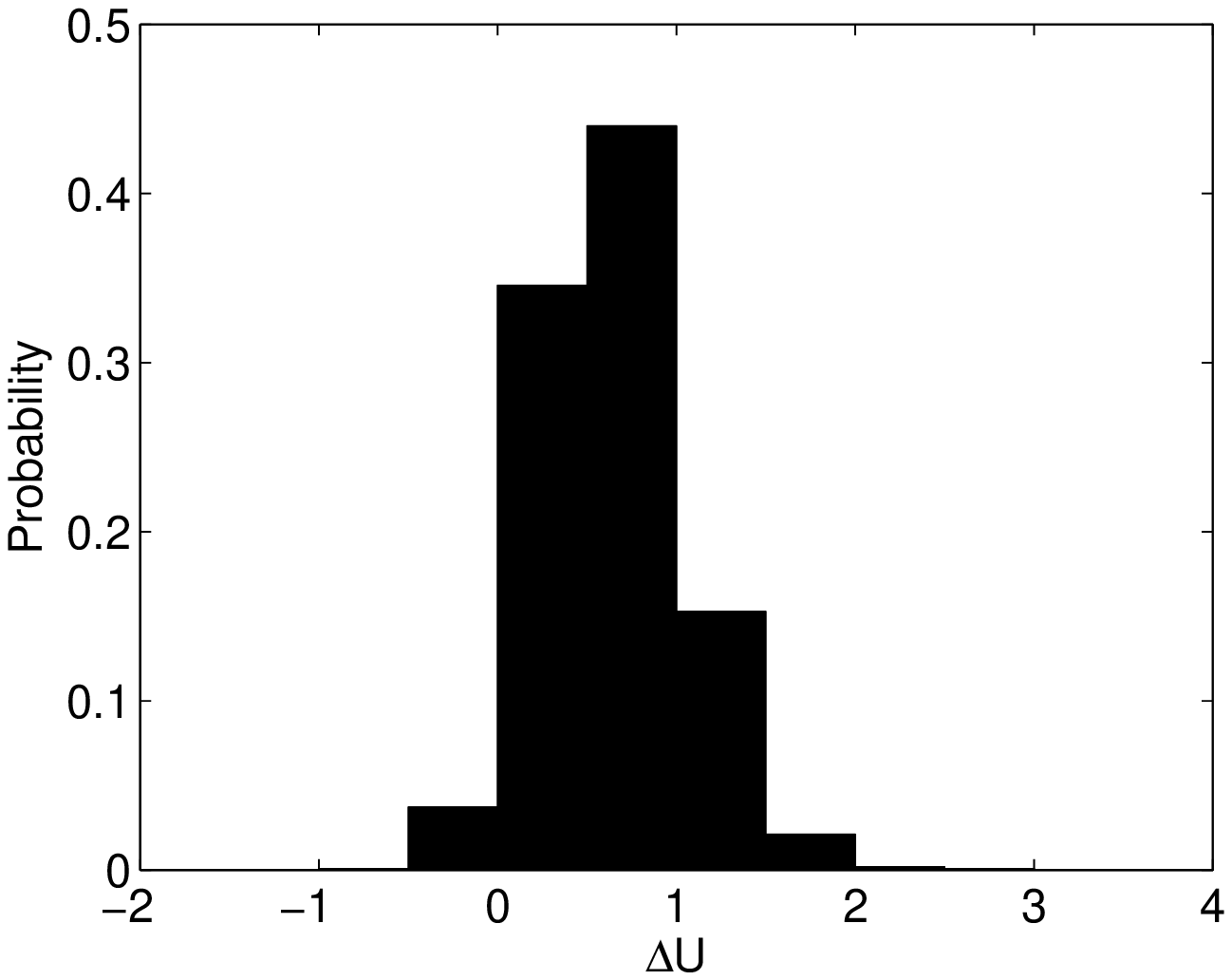}
\label{fig:dU_hist}
}
\subfigure[]{
\includegraphics[width=4in]{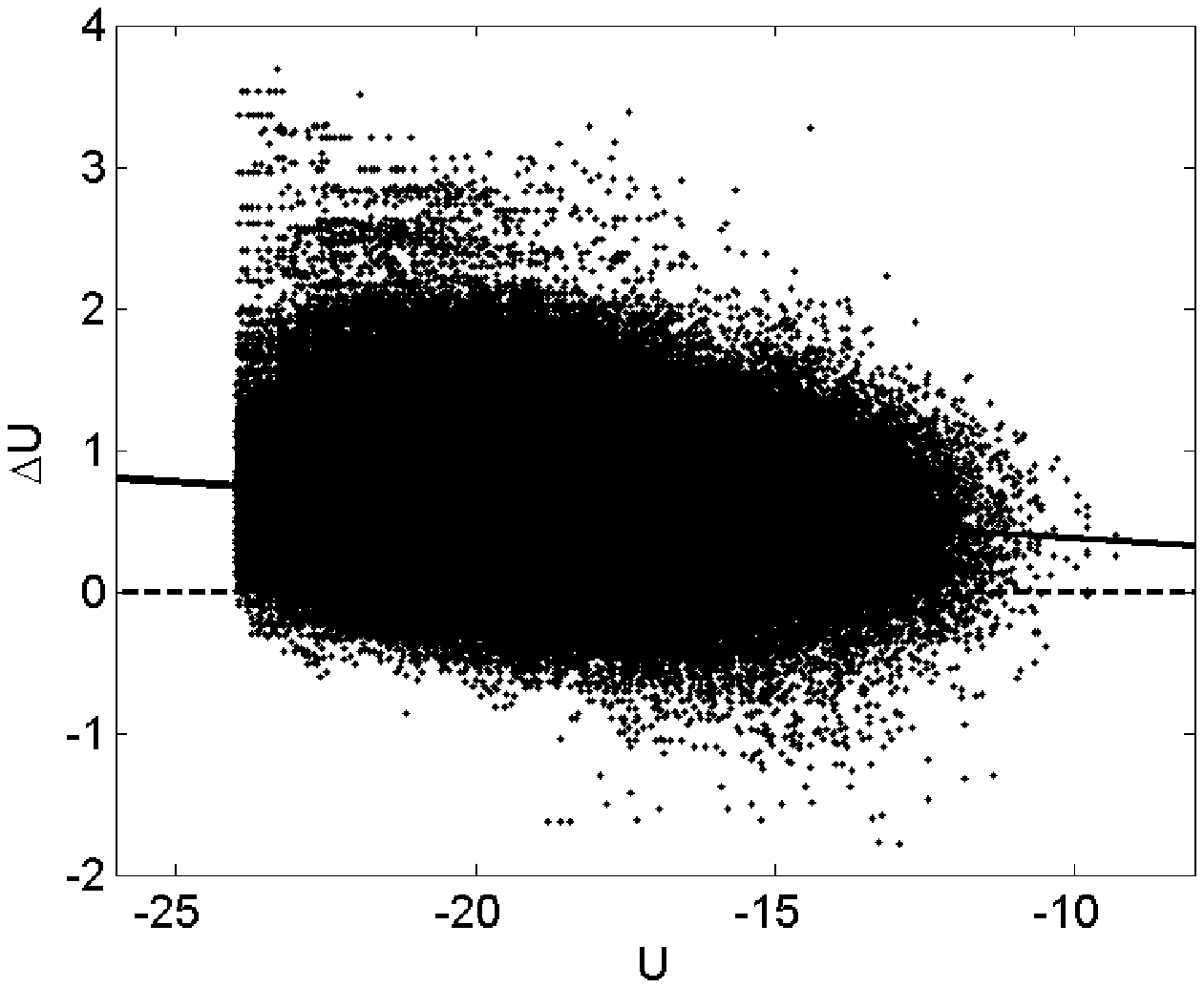}
\label{fig:dU_vs_U}
}
\end{center}
\caption{\subref{fig:dU_hist} The histogram of the energy difference $\Delta U = \hat{U} - U$ associated with a point mutation in the generalized $NK$ model, in which $U$ and $\hat{U}$ are the energy values before and after the point mutation, respectively. The values of $U$ and $\hat{U}$ are calculated for 1000 B cells at 61 generations. The values of $\Delta U$ ranged from $-1.78$ to $3.69$. The histogram was equally divided in the interval $\left(-2, 4\right)$ by 12 bins. The probability of the mutations with $\Delta U < 0$ is 0.038. \subref{fig:dU_vs_U} The original energy, $U$, versus the energy difference, $\Delta U$, during 20 runs of the generalized $NK$ model. The horizontal dashed line is $\Delta U = 0$. The solid line is the trend line between $U$ and $\Delta U$ fit through the data. On average $U$ decreases with the generation. At generation 0, $U$ is typically close to $-10$, and at generation 60, $U$ is typically close to $-25$.}
\label{fig:dU}
\end{figure}

The generalized $NK$ model used in this work was used to compute the distribution of  energy $U$ and $\Delta U$, the energy difference caused by a point mutation. At each generation from 0 to 60, $\Delta U$ was calculated by  introducing a random mutation in each B cell.  Figure \ref{fig:dU}\subref{fig:dU_hist} describes the distribution of $\Delta U$. The point mutations with $\Delta U < 0$ comprise 3.8\% of the distribution. Experimentally, this fraction is observed to be 4.9\% of the distribution \cite{Zhang2010}. Figure \ref{fig:dU}\subref{fig:dU_vs_U} plots the data points $\left(U, \Delta U\right)$ generated in the simulation in a two-dimensional space. The linear regression between $U$ and $\Delta U$ produces a trend line:
\begin{equation}\label{eq:U_dU}
\Delta U = -0.027 U + 0.109.
\end{equation}
The slope of this trend line is significantly different from zero ($p\mbox{-value} < 2.2 \times 10^{-16}$). The negative slope is expected because $\Delta U$ is expected to have a symmetric distribution with center zero when $U$ equals to zero but to be skewed for negative $U$ in the generalized $NK$ model \cite{Bogarad1999,Deem2003}. Equation \ref{eq:U_dU} is consistent with this expectation within error bars. Figure \ref{fig:dU}\subref{fig:dU_vs_U} shows that the correlation between $U$ and $\Delta U$ is weak ($R^2 = 0.018$). Weak correlation between $U$ and $\Delta U$ is observed in the PINT database of experimental data of affinity-related amino acid mutations  \cite{Zhang2010}.

\subsection{Immune response to multiple antigens}
\label{sec:ode}

\begin{figure}[!ht]
\begin{center}
\includegraphics[width=4in]{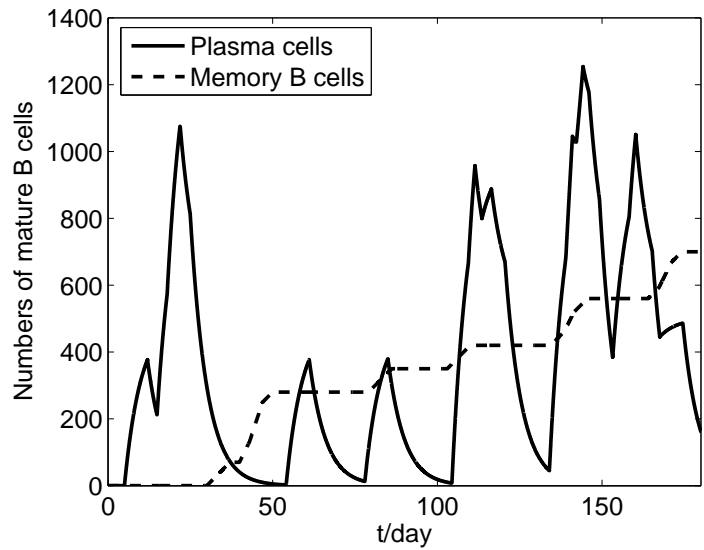}
\end{center}
\caption{Dynamics of mature B cells on days 0--180 in one zebrafish reacting to all types of the antigen in the environment. The numbers of plasma cells and memory B cells are the sums of $x_i$ and $y_i$, respectively. The inoculation time of the antigens followed a Poisson process and the identity of which is randomly chosen, as described in the main text. The dynamics shown in this figure are from one trajectory of the Poisson challenge process.}
\label{fig:B_trajectory}
\end{figure}

In the aquatic environment of the experiment, the fish are challenged by multiple antigens over their lifespan. Each type of antigen induces a distinct immune response and produces different mature B cells.
The solution of the ODE model defined by equations \ref{eq:ode_1} and \ref{eq:ode_2} is determined by the presence of each antigen $i$ denoted by $v_i\left(t\right)$. In the simulation, we first randomly selected antigens to which the zebrafish were exposed at several time points, from a diversity of 10 total antigens.  As discussed in Section \ref{sec:ode_model}, the rate of encountering antigen is 1/30 day$^{-1}$.
 Figure \ref{fig:B_trajectory} shows the dynamics of the total plasma cells and the total memory B cells in one run of the simulation. Each peak of plasma B cells corresponds to an antigen inoculation. In the particular instance shown in Figure \ref{fig:B_trajectory}, the zebrafish is stimulated 
12 times with 8 distinct antigens, leading to 8 primary responses and 
4 secondary responses
 in its six-month life span. The first innoculation with antigen induces a primary immune response response. Subsequent innoculations with antigens induce a primary immune response if it is a new antigen or a secondary immune response if the zebrafish was previously infected by an antigen of that type.  The VDJ recombinants are assigned in the primary response and remain the same in any secondary response to the same antigen $i$.

\begin{figure}[!ht]
\begin{center}
\subfigure[]{
\includegraphics[width=2.5in]{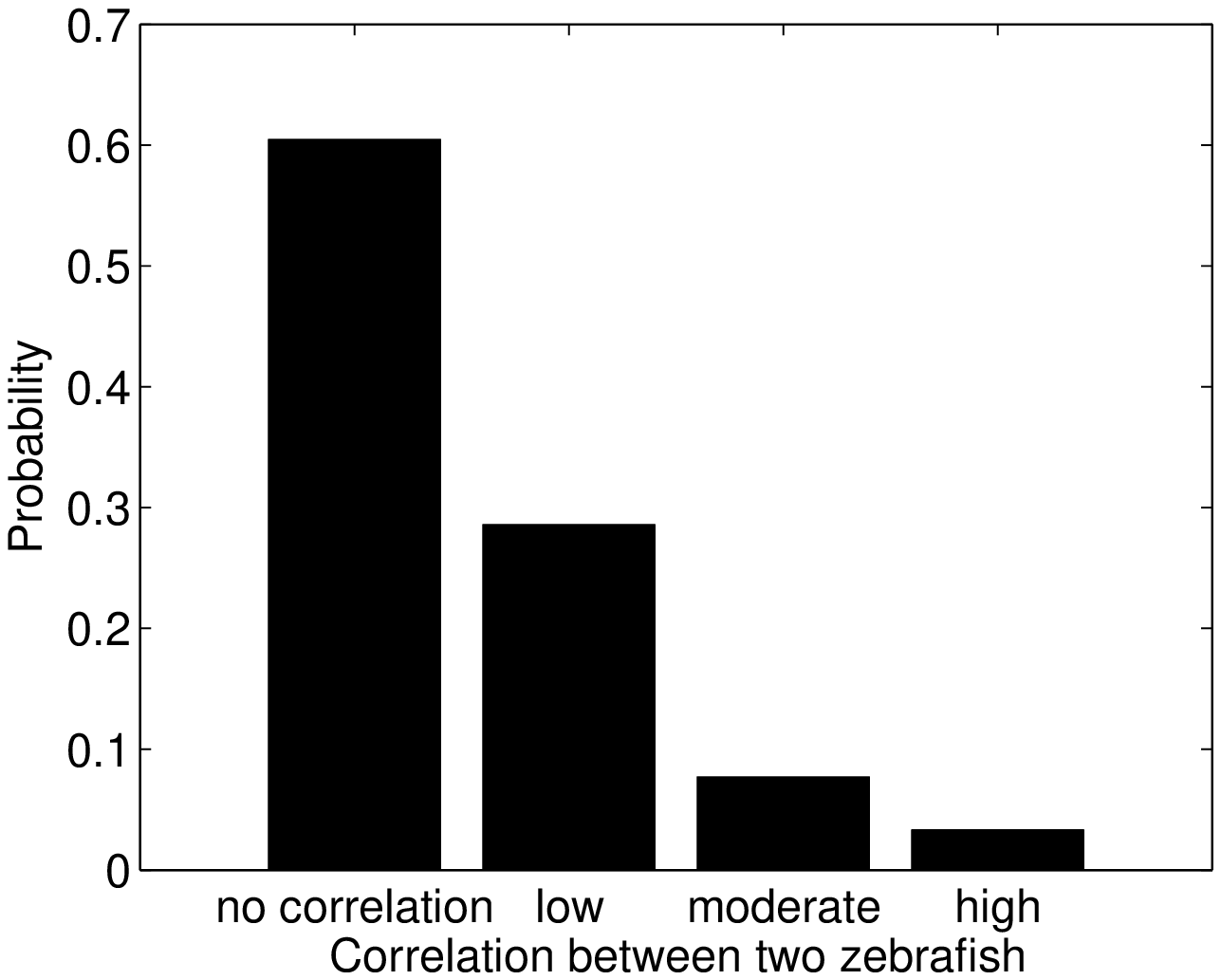}
\label{fig:corr_experiment}
}
\subfigure[]{
\includegraphics[width=2.5in]{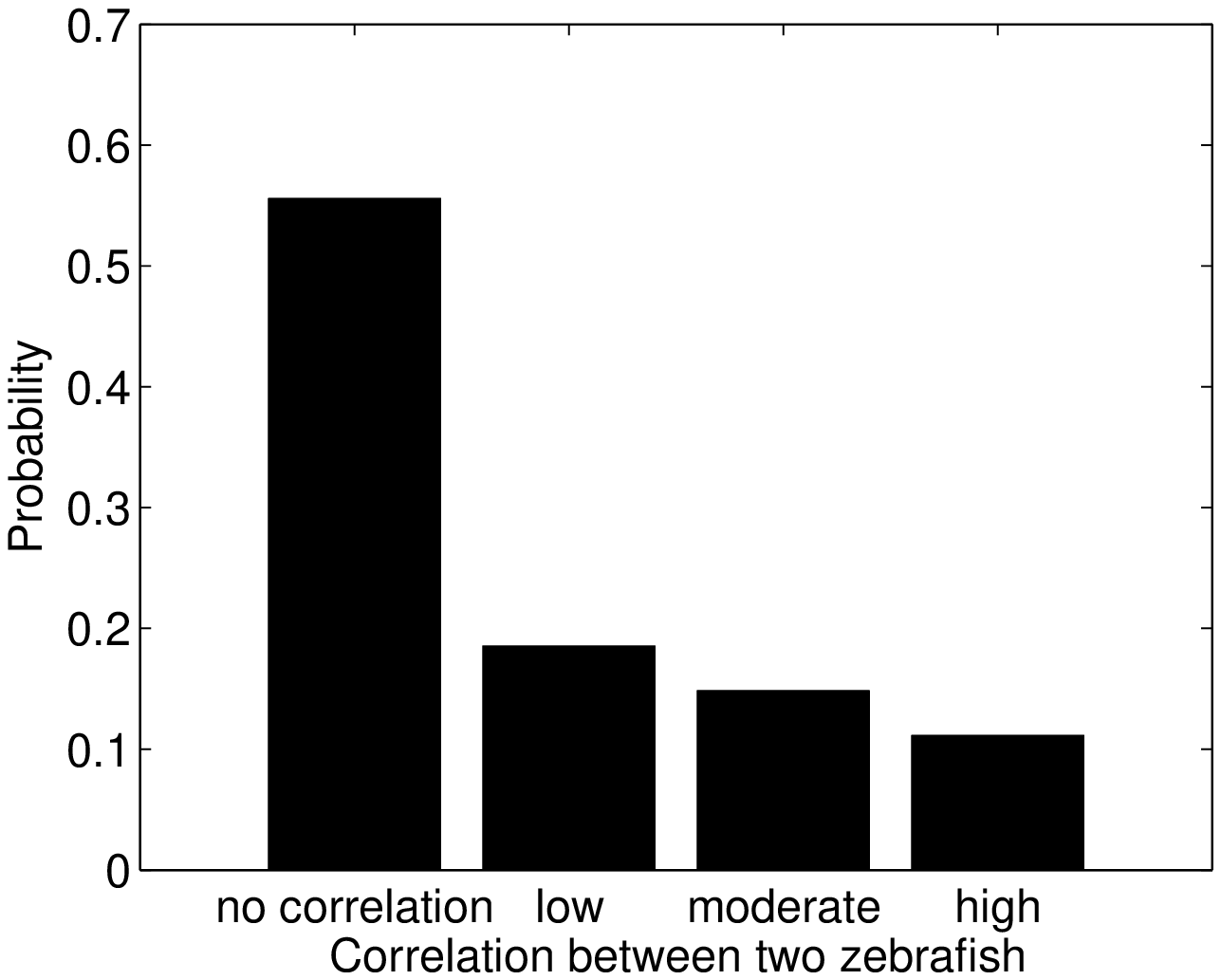}
\label{fig:corr_experiment2}
}
\subfigure[]{
\includegraphics[width=2.5in]{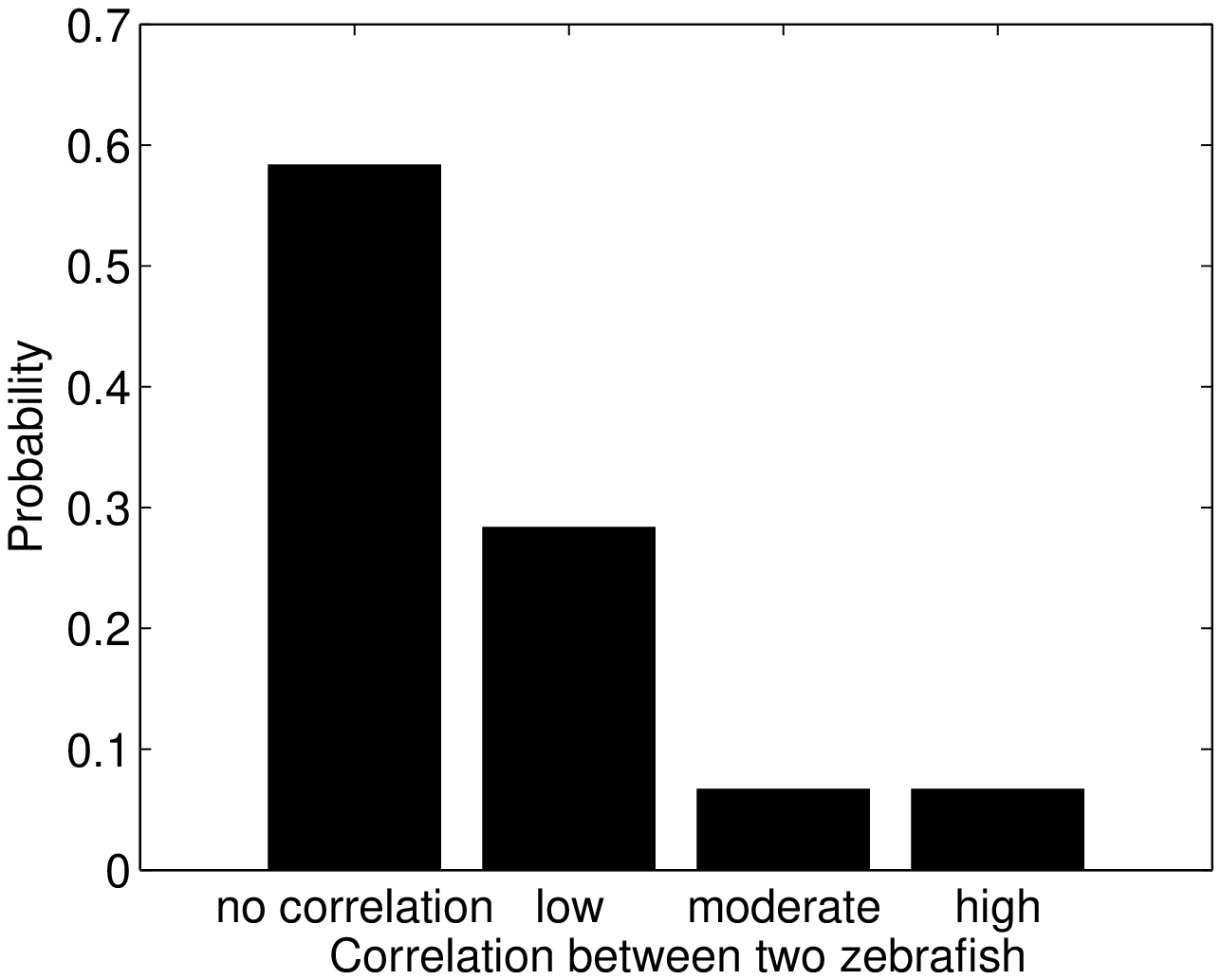}
\label{fig:corr_model}
}
\end{center}
\caption{Distribution of correlation coefficients between the VDJ usage in B cell repertoires in distinct zebrafish on day 180. Using the categorization scheme in \cite{Weinstein2009}, the correlation coefficients $r$ fall into four bins: no correlation with $r < 0.1$, low correlation with $0.1 \le r < 0.2$, moderate correlation with $0.2 \le r < 0.5$, and high correlation with $r \le 0.5$. The height of each bar equals the probability of the correlation coefficient being in each bin. \subref{fig:corr_experiment} Correlation coefficient data between all the zebrafish in the experiment \cite{Weinstein2009}. \subref{fig:corr_experiment2} Correlation coefficient data between the zebrafish living in the same aquarium in the experiment \cite{Weinstein2009}. \subref{fig:corr_model} A total of 6000 correlation coefficients were generated by the model.}
\label{fig:corr_histogram}
\end{figure}

There is substantial similarity between the predicted distribution of the correlation coefficients of VDJ usages in pairs of fish and that measured  in experiment \cite{Weinstein2009}. We adopted the experimental categorization scheme for the correlation coefficient \cite{Weinstein2009}, which classifies the correlation coefficients $r$ in four bins: no correlation with $r < 0.1$, low correlation with $0.1 \le r < 0.2$, moderate correlation with $0.2 \le r < 0.5$, and high correlation with $r \ge 0.5$. The 14 zebrafish belonged to four different families, and each family lived in a separate aquarium. There are $14 \times 13 / 2 = 91$ correlation coefficients between all pairs of the 14 zebrafish. Figure \ref{fig:corr_histogram}\subref{fig:corr_experiment} shows the relative frequency of all 91 correlation coefficients in experiment in each of the four bins.  We also calculate the 27 correlation coefficients between the VDJ usages in the zebrafish from the same family. Figure \ref{fig:corr_histogram}\subref{fig:corr_experiment2} shows the relative frequency of these 27 correlation coefficients in experiment in each of the four bins. Figure \ref{fig:corr_histogram}\subref{fig:corr_model} shows the distribution of correlation coefficients from the simulation. The predicted and measured histograms are similar: Figure \ref{fig:corr_histogram}\subref{fig:corr_experiment} and Figure \ref{fig:corr_histogram}\subref{fig:corr_model} ($p\mbox{-value} = 0.62$, Pearson $\chi^2$ test) and  Figure \ref{fig:corr_histogram}\subref{fig:corr_experiment2} and Figure \ref{fig:corr_histogram}\subref{fig:corr_model} ($p\mbox{-value} = 0.22$, Pearson $\chi^2$ test).

The prediction relies on the value of $p$, calculated from first principles from the generalized $NK$ model.
No adjustment of the generalized $NK$ model from prior literature form or parameter values was done. The generalized $NK$ model is well established based on the experimental data in immunology \cite{Janeway2005} and has successfully simulated the immune responses to influenza \cite{Deem2003} and peptides \cite{Park2004}. In this study, we do not arbitrarily change the form of the generalized $NK$ model or add any assumption. So $p = 0.327$ is a first-principles calculation. We make two assumptions on the experimental data: (1) there are $N_\mathrm{ag} = 10$ distinct types of antigen in the environment, and (2) the average time span between two events of infection is $\lambda = 30$ days. These two assumptions are necessary because we do not have information on the antigens infecting the zebrafish in experiment \cite{Weinstein2009}.

\begin{figure}[!ht]
\begin{center}
\subfigure[]{
\includegraphics[width=2.5in]{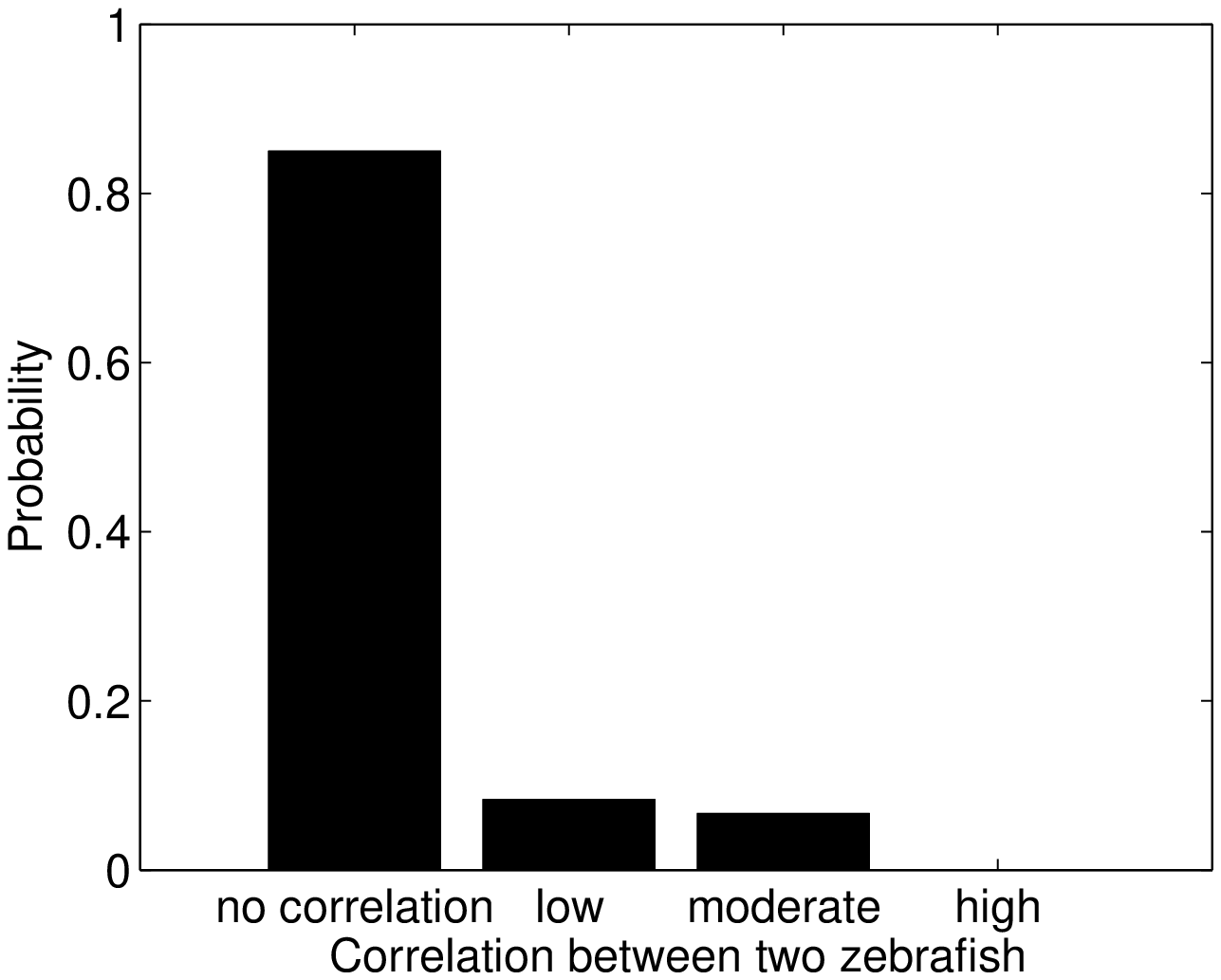}
\label{fig:paramscan_T_90}
}
\subfigure[]{
\includegraphics[width=2.5in]{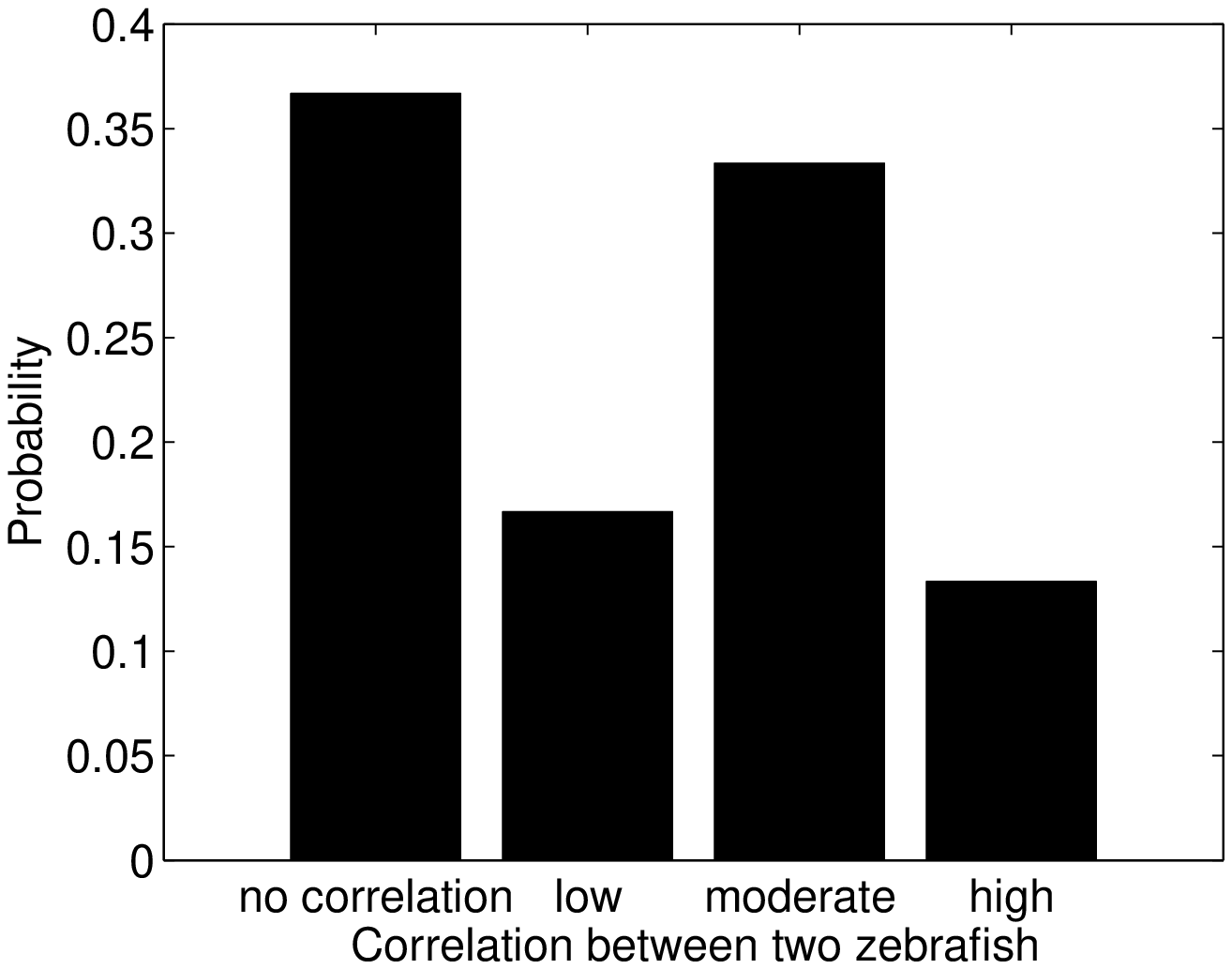}
\label{fig:paramscan_T_360}
}
\subfigure[]{
\includegraphics[width=2.5in]{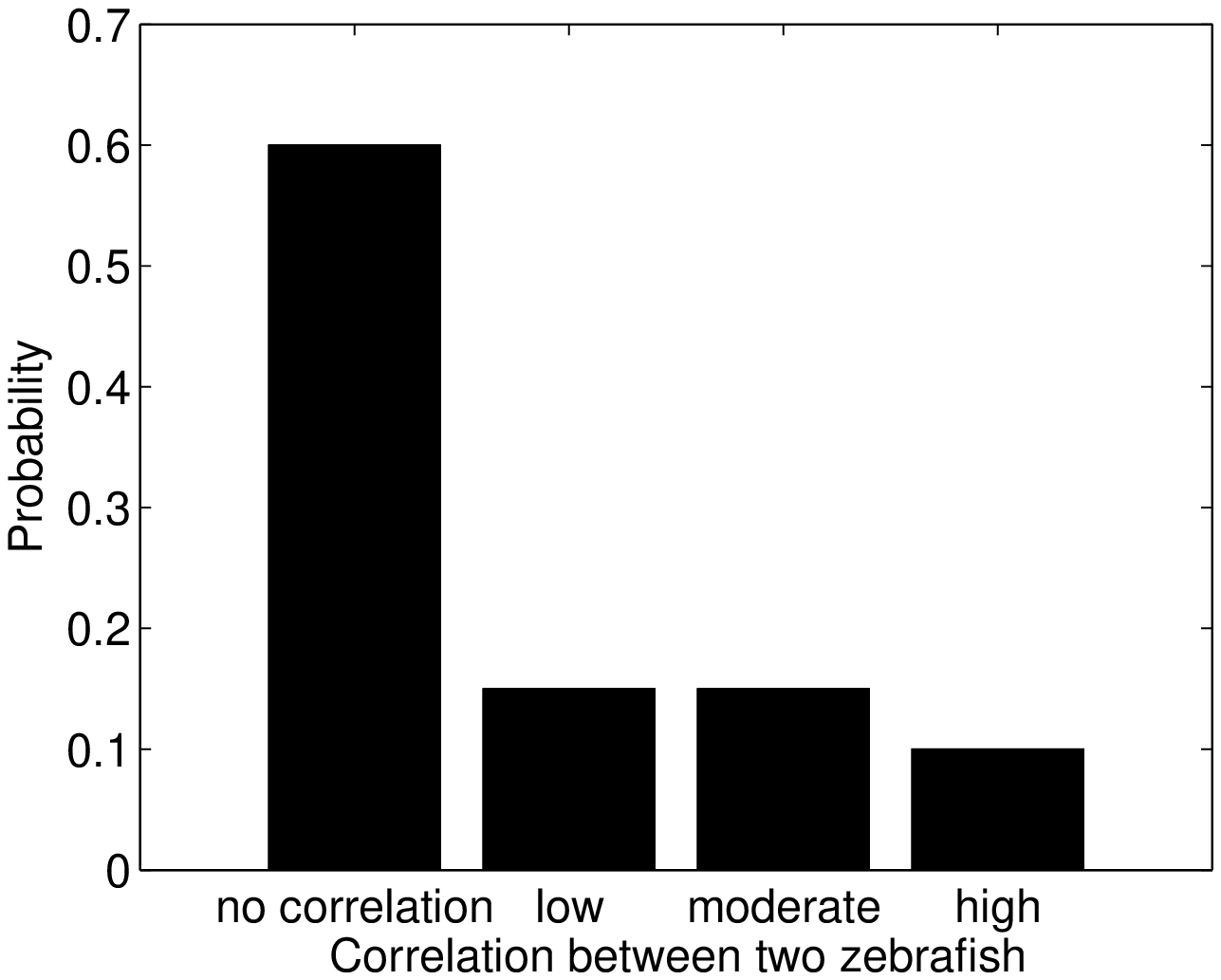}
\label{fig:paramscan_N_AG_5}
}
\subfigure[]{
\includegraphics[width=2.5in]{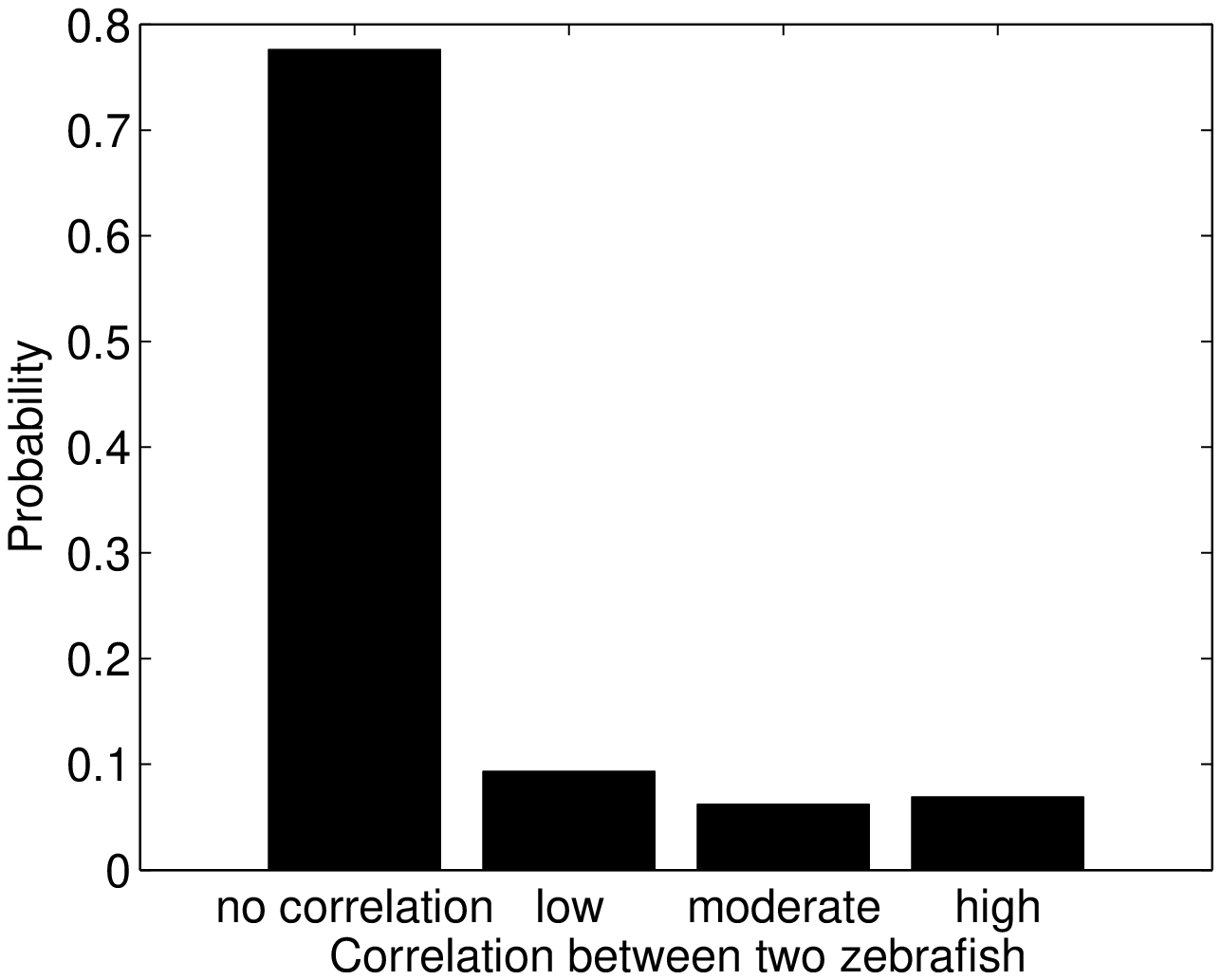}
\label{fig:paramscan_N_AG_20}
}
\subfigure[]{
\includegraphics[width=2.5in]{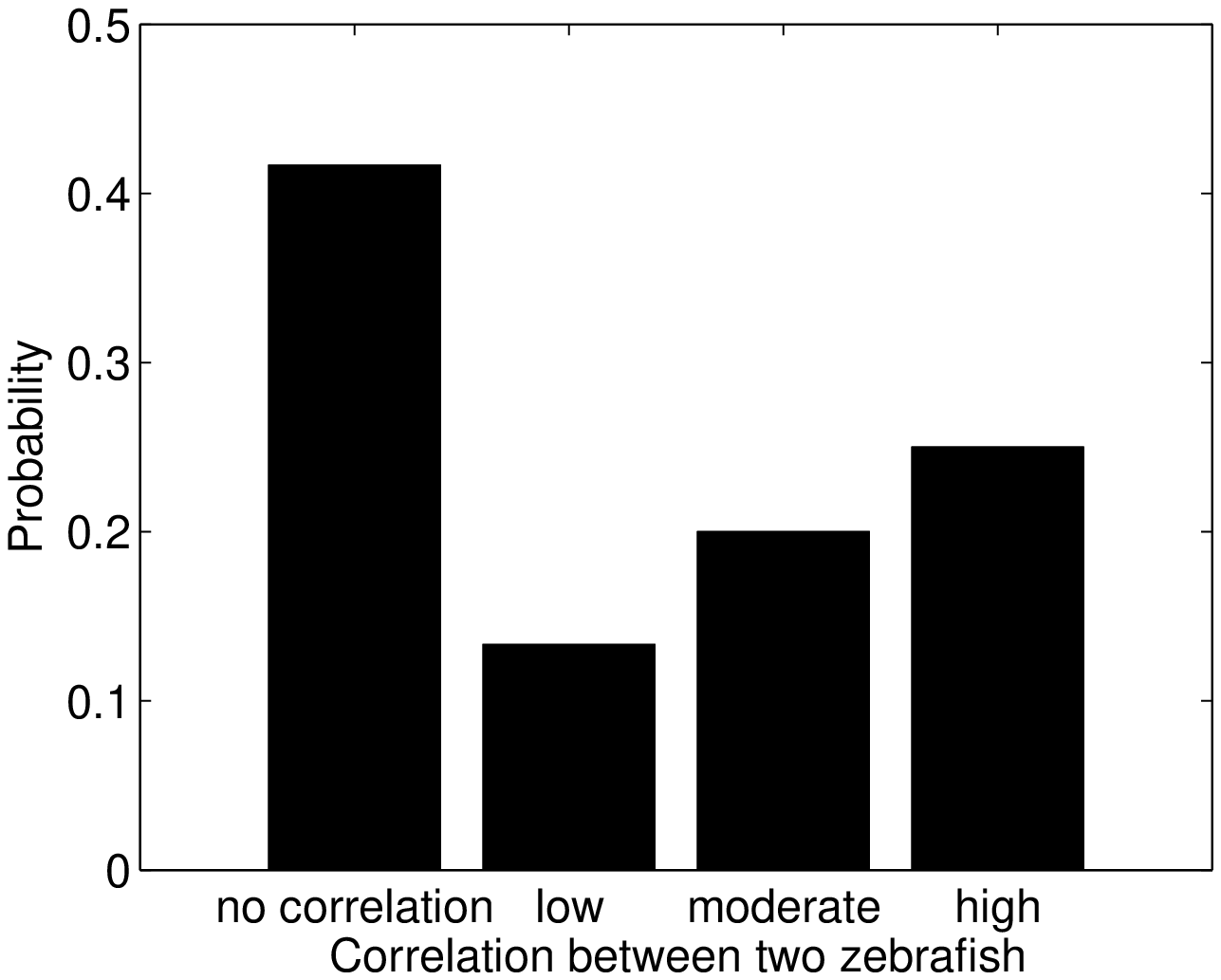}
\label{fig:paramscan_lambda_15}
}
\subfigure[]{
\includegraphics[width=2.5in]{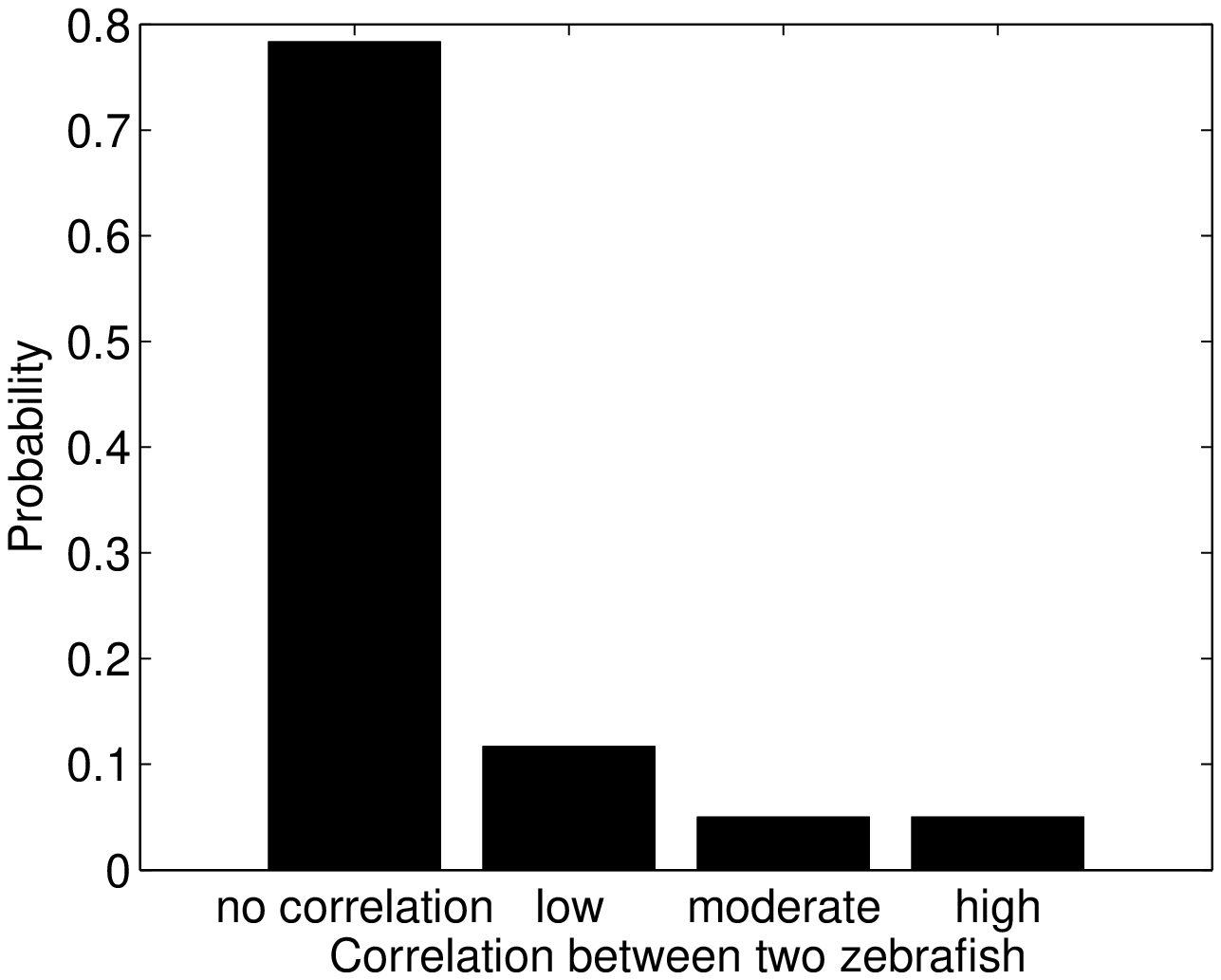}
\label{fig:paramscan_lambda_60}
}
\end{center}
\caption{One-way sensitivity analysis of the correlation histogram in Figure \ref{fig:corr_histogram}\subref{fig:corr_model}. The length of the experiment was changed from 180 days to 90 days \subref{fig:paramscan_T_90} and to 360 days \subref{fig:paramscan_T_360}. The number of antigen types was changed from 10 to 5 \subref{fig:paramscan_N_AG_5} and to 20 \subref{fig:paramscan_N_AG_20}. The average time span between antigen infections was changed from 30 days to 15 days \subref{fig:paramscan_lambda_15} and to 60 days \subref{fig:paramscan_lambda_60}.}
\label{fig:paramscan_histogram}
\end{figure}

We performed a one-way analysis to test the sensitivity of the correlation histogram to the length of the experiment, the number of antigen types, and the frequency of antigen infection.  Figure \ref{fig:paramscan_histogram} shows the results. First, we find that increasing the length of experiment leads to higher correlation in VDJ usage in a pair of zebrafish, as shown in Figures \ref{fig:paramscan_histogram}\subref{fig:paramscan_T_90} and \subref{fig:paramscan_T_360}. 
Selection in the immune system is what causes the correlation, and a longer experiment allows for more selection.
Second, Figures \ref{fig:paramscan_histogram}\subref{fig:paramscan_N_AG_5} and \subref{fig:paramscan_N_AG_20} indicate that increasing the number of antigen types slightly decreases the correlation in VDJ usage.  This is presumably because a greater antigen diversity leads to greater primary responses and random VDJ assignments rather than secondary responses, in which the VDJ usage remains unchanged from the primary response.
 Third, Figures \ref{fig:paramscan_histogram}\subref{fig:paramscan_lambda_15} and \subref{fig:paramscan_lambda_60} indicate that increasing the frequency of antigen infection results in higher correlation in VDJ usage.
This is because the selection that causes correlation is initiated only by exposure to antigen.
 Both a longer experiment and a higher frequency of antigen infection cause more primary and secondary immune responses in the fish. The sensitivity analysis in Figure \ref{fig:paramscan_histogram} implies that repeated antigen exposure gradually localizes the VDJ usage in a pair of fish to the same region.
%KEYAO ADD

\section{Conclusion and outlook}

In this work, a multi-scale model is developed to describe the diversity of VDJ recombinations in zebrafish exposed to a diversity of antigens over their lifespan.
The multi-scale model comprises a delay ODE model and a generalized $NK$ model. The delay ODE model is a mean-field approximation averaging over the genotypes of B cells specific to one type of antigen. The dynamical system defined by the ODEs describes the average B cell response to  the challenge of antigen infection. The solution of the ODE system gives the composition of the mature B cell repertoire at any time point between 0 and 180 days. The correlation between VDJ usage in the B cell repertoires in a pair of zebrafish is  calculated.
 The predicted correlation  of VDJ usage, Figure  \ref{fig:corr_histogram},  is similar to that from the experiments \cite{Weinstein2009}, and predictions for different experimental conditions are presented, Figure \ref{fig:paramscan_histogram}.

The generalized $NK$ model describes evolution of B cells specific to a single type of antigen. Each type of antigen in the environment leads to distinct memory B cell evolution.  Selection over B cells removes genotypes with high energy and hence substantially localizes the B cells in the rugged energy landscape and in the sequence and  VDJ recombination space. The localization, however, is not a deterministic process. The primary immune response has a probability $p$ to localize to the same VDJ recombination in a pair of zebrafish, and this $p$ is calculated by the generalized $NK$ model. The probability $p$ increases with the number of B cells reacting to a specific type of antigen, while decreasing with the somatic hypermutation rate and the survival rate in the selection. A higher mutation rate increases the uncertainty in the B cell maturation process. A high level of selection in the immune system  increases the tendency to fix the B cell in the current location in the energy landscape. We found the probability p to decrease with  mutation rate and increase with selection intensity.

The correlation between VDJ usage in a pair of zebrafish shows that evolution in the immune system is not a completely random process.  Somatic evolution is not an unbiased random walk.  The finite VDJ repertoire size of the fish mean that a pair of fish have a nonzero probability, $p$, to localize upon the same VDJ recombination.
 In the ODE model, correlation of the VDJ usage in the B cell repertoire on day 180 also shows that the B cell somatic evolution in a pair zebrafish may be correlated if they are challenged by the same antigen at the same time. Consequently, closely related antigen challenges may induce immune responses with similar characters in distinct individuals.  Longer expeirments or more frequent antigen exposure significantly increases the correlation between fish, while reduced antigen diversity in the environment more modestly increases this correlation between fish.

This multi-scale model is flexible and extensible to analyze the immune system dynamics in zebrafish under different conditions or in other species. This study on the zebrafish B cell repertoire can be extended by analyzing the immune system in zebrafish at different ages. The B cell repertoire in these individuals may reveal the dynamics of development of immune systems challenged by various types of antigen, and allow for further tests of the model. Study of VDJ correlations in fish with closely related genomes may be interesting. We propose sequencing of the B cell repertoire of zebrafish in different families living in different controllable environments to test the contribution to VDJ usage of two factors: zebrafish genome and antigens in the environment. If the correlation of VDJ usage in distinct zebrafish strongly depend on their genetic similarity in other than by obvious dependence on the VDJ region, their genomes are a determinant of VDJ usage. If not, VDJ usage is independent of the genome of each individual.

\section*{Acknowledgments}

This work was partially supported by DARPA grant HR 00110910055.

\section*{References}

\bibliographystyle{unsrt}
\bibliography{references}

\section*{Table and figures}

\begin{table}[!ht]
\caption{Comparison and connection between the delay ODE model and the generalized $NK$ model. The delay ODE model describes the dynamics of antigen $i$ specific B cells. The generalized $NK$ model describes the amino acid sequences,  the binding affinity to the antigen, and the hypermutation and selection of B cells. The output of the generalized $NK$ model, $p$, is the input to the delay ODE model.}
\vspace{\baselineskip}
%\begin{indented}
{\footnotesize
\begin{tabular}{|l|l|l|}\hline
                         & {\bf The delay ODE model} & {\bf The generalized $NK$ model} \\\hline\hline
Number of zebrafish used & 1                   & 2 \\\hline
Type of antigens         & Multiple            & 1 \\\hline
\begin{minipage}[t]{0.3\textwidth}Type of antigen-specific mature B cells\end{minipage} & Multiple            & 1 \\\hline
\begin{minipage}[t]{0.3\textwidth}Number of primary/secondary immune responses simulated\end{minipage} & \begin{minipage}[t]{0.4\textwidth}Multiple (primary)\\Multiple (secondary)\end{minipage} & \begin{minipage}[t]{0.3\textwidth}1 (primary) \\ 1 (secondary)\end{minipage} \\\hline
Simulation length        & \begin{minipage}[t]{0.4\textwidth}180 days, as in the experiment \cite{Weinstein2009}\end{minipage} & \begin{minipage}[t]{0.3\textwidth}10 days for the primary immune response, 10 days for the secondary immune response\end{minipage} \\\hline
\begin{minipage}[t]{0.3\textwidth}Information on the mature B cells given by the model\end{minipage} & \begin{minipage}[t]{0.4\textwidth}Numbers of plasma cells and memory B cells\end{minipage} &
\begin{minipage}[t]{0.3\textwidth}Number of mature B cell (including plasma cells and memory B cells), amino acid sequence of each B cell, binding affinity to the certain antigen of each B cell\end{minipage} \\\hline
\begin{minipage}[t]{0.3\textwidth}Minimal component in the model\end{minipage} & B cell &
\begin{minipage}[t]{0.3\textwidth}Amino acid residue in the V region of the antibody\end{minipage} \\\hline
Mathematical basis & Delay ODE & Spin-glass theory \cite{Bogarad1999,Deem2003} \\\hline
Input & \begin{minipage}[t]{0.4\textwidth}(1) A series of antigens, each of which infect the zebrafish in one or more time points \\(2) The probability $p$ that the mature B cells in two zebrafish have the same VDJ recombination \end{minipage} & None \\\hline
Output & \begin{minipage}[t]{0.4\textwidth}Dynamics of plasma cells and memory B cells specific to each type of antigen\end{minipage} & \begin{minipage}[t]{0.3\textwidth}The probability $p$\end{minipage} \\\hline
Connection of the two models &
\begin{minipage}[t]{0.4\textwidth}(1) For each type of antigen $i$, gives the number of plasma cells $N_{\mathrm{p}i}$ and memory B cells $N_{\mathrm{m}i}$ specific to any antigen $i$ at any time point \\(2) Assigns identical (probability: $p$) or distinct (probability: $1 - p$) VDJ recombination to ($N_{\mathrm{p}i} + N_{\mathrm{m}i}$) mature B cells specific to antigen $i$ in either zebrafish.\end{minipage} &
\begin{minipage}[t]{0.3\textwidth}For each type of antigen $i$, gives the probability $p$\end{minipage} \\\hline
\end{tabular}
}
%\end{indented}
\label{tab:model}
\end{table}

% Source: Zebrafish\Figures\scheme\raw_cut (MS Visio file)

% Notebook 4, P.76
% Source code: king.rice.edu:/scratch8/kpan/gnk_8/size_2000 and size_2000_rank
% C:\Users\Keyao\Documents\Influenza\ML\VDJ\random_energy_1\localization\n_seq_vdj.m

% Source code: from king.rice.edu:/scratch8/kpan/gnk_5/size_2000/
% C:\Users\Keyao\Documents\Influenza\ML\VDJ\random_energy_1\traj\prod_1.m

% Figure (a)
% Notebook 4, P.74
% Source code: from king.rice.edu:/scratch8/kpan/gnk_5
% C:\Users\Keyao\Documents\Influenza\ML\VDJ\random_energy_1\proc_5\R2_size.m
%
% Figure (b)
% Source code: from king.rice.edu:/scratch8/kpan/gnk_8, fort.15 -> n_seq_1000/2000/10000.dat, fort.17 -> n_vdj_1000/2000/10000.dat
% C:\Users\Keyao\Documents\Influenza\ML\VDJ\random_energy_1\localization\n_vdj2\num_vdj.m

% Notebook 4, P.76
% Source code: sugar.rice.edu:/users/kp3/VDJ/param_scan/proc_1
% C:\Users\Keyao\Documents\Influenza\ML\VDJ\random_energy_1\proc_5\param_scan\proc_1.m

% Notebook 4, P.76
% Source code: sugar.rice.edu:/users/kp3/VDJ/param_scan/proc_1
% C:\Users\Keyao\Documents\Influenza\ML\VDJ\random_energy_1\proc_5\param_scan\proc_1.m

% Notebook 4, P.75
% Source code: king.rice.edu:/scratch8/kpan/gnk_U_dU/
% C:\Users\Keyao\Documents\Influenza\ML\VDJ\random_energy_1\proc_5\U_dU\U_dU.m

% Notebook 4, P.78
% Source code: sugar.rice.edu:/users/kp3/VDJ/ODE/proc_1/solver_1_3a.m
% C:\Users\Keyao\Documents\Influenza\ML\VDJ\ode\proc_4\solver_1_3a.m

% Notebook 4, P.78
% Source code: sugar.rice.edu:/users/kp3/VDJ/ODE/proc_1/solver_1_3a.m, submit.pbs
% C:\Users\Keyao\Documents\Influenza\ML\VDJ\ode\stat_test2\chi2_stat1.m

% Source code: sugar.rice.edu:/users/kp3/VDJ/ODE/proc_1_paramscan, in six directories
% C:\Users\Keyao\Documents\Influenza\ML\VDJ\ode\stat_test3, in six directories

\end{document}